%% file: main.tex
\documentclass[twocolumn]{aastex631}
\usepackage{multirow}
\usepackage{xspace}
\usepackage{soul}

\newcommand\obsid{{\sc obsid}}
\newcommand\pcube{\texttt{PCUBE}}
\newcommand\pdx{${\rm PD}_{\rm X}$}
\newcommand\pdo{${\rm PD}_{\rm Opt}$}
\newcommand\pdra{${\rm PD}_{\rm radio}$}
\newcommand\pax{${\rm PA}_{\rm X}$}
\newcommand\pao{${\rm PA}_{\rm Opt}$}
\newcommand\para{${\rm PA}_{\rm Radio}$}
\newcommand\mdp{${\rm MDP}_{99}$}
\newcommand{\fluxcgs}{\ensuremath{\mathrm{erg}\,\mathrm{s}^{-1}\,\mathrm{cm}^{-2}}}
\counterwithout{figure}{section}

\begin{document}

\title{X-ray and multiwavelength polarization of Mrk~501 from 2022 to 2023}

\include{authorlist.tex}

%\collaboration{142}{(the IXPE Collaboration)}

\begin{abstract}    
We present multiwavelength polarization measurements of the luminous blazar Mrk~501 over a 14-month period. The 2--8~keV X-ray polarization was measured with the Imaging X-ray Polarimetry Explorer (IXPE) with six 100-ks observations spanning from 2022 March to 2023 April. Each IXPE observation was accompanied by simultaneous X-ray data from NuSTAR, Swift/XRT, and/or XMM-Newton. Complementary optical-infrared polarization measurements were also available in the B, V, R, I, and J bands, as were radio polarization measurements from 4.85 GHz to 225.5 GHz. Among the first five IXPE observations, we did not find significant variability in the X-ray polarization degree and angle with IXPE. However, the most recent sixth observation found an elevated polarization degree at $>3\sigma$ above the average of the other five observations. The optical and radio measurements show no apparent correlations with the X-ray polarization properties. Throughout the six IXPE observations, the X-ray polarization degree remained higher than, or similar to, the R-band optical polarization degree, which remained higher than the radio value. This is consistent with the energy-stratified shock scenario proposed to explain the first two IXPE observations, in which the polarized X-ray, optical, and radio emission arises from different regions.

\end{abstract}
\keywords{}

\section{Introduction} \label{sec:intro}

\begin{table*}
\footnotesize
\begin{tabular}{cccccc}
\hline\hline
IXPE \obsid\ & Date & Exp (ks) &  Net count rate (cts/s) & MDP$_{99}$ (\%) & Other X-ray data$^\star$ \\
\hline
01004501 (IXPE1) & 2022-03-07 & 104& 0.24 &  6.4& NuSTAR 60701032002 (20 ks) \\
01004601 (IXPE2) & 2022-03-27 & 87 & 0.44 &  5.0 & NuSTAR 60702062004 (20 ks)\\
01004701 (IXPE3) & 2022-07-09 & 98 & 0.21 &  7.5 & Swift/XRT 00011184$^{\star\star}$ (0.9 ks)\\
02004601 (IXPE4) & 2023-02-12 & 95 &  0.12 &  9.5 & XMM 0902111901 (14 ks) \\
02004501 (IXPE5) & 2023-03-19 & 102&  0.14 &  8.4 & XMM 0902112201 (9 ks)\\
02004701 (IXPE6) & 2023-04-16 & 94 &  0.15  & 8.5& Swift/XRT 00015411 (1.1 ks) \\
\hline
\end{tabular}
\caption{Details of each IXPE observation studied in this work. The net count rates shown here are the average background subtracted 2--8~keV count rate per second. The Minimum Detectable Polarization at 99\% confidence,  MDP$_{99}$, is defined in  Sec.~\ref{sec:pcube}.\\ 
$^\star$All IXPE observations reported here also have Swift XRT and UVOT coverage.\\
$^{\star\star}$ XMM observed Mrk 501 simultaneously with IXPE, but XMM \obsid\ 0902110701 suffered from significant background flaring and was not included in this analysis.
}\label{tab:obs}
\end{table*}

Blazars are a class of radio-loud active galactic nuclei (AGN) featuring relativistic jets powered by 
accreting supermassive black holes, with one of the jets pointing within $\sim10\degr$ of our line of sight \citep[e.g.,][]{Hovatta2019}. The strongly beamed jet emission dominates the spectral energy distribution (SED), making blazars prime targets for studying particle acceleration and non-thermal emission mechanisms that occur in the jets \citep[see, e.g.,][]{Blandford2019}. 
%\citep[see, e.g.,][]{Blandford2019,Boettcher2019,Boettcher2022}.

The advent of the Imaging X-ray Polarimetry Explorer \citep[IXPE;][]{Weisskopf2022} has enabled X-ray polarization measurements of blazars, which inform us about the magnetic field geometry during periods of efficient particle acceleration in the jets \citep{Zhang2013,Zhang2019,Liodakis2019,Peirson2022}. The first IXPE observation of a blazar confirmed Mrk 501 to be linearly polarized at photon energies of 2--8 keV, with a polarization degree (hereafter PD) of $\sim 10\%$, several times the concurrent values at optical and radio wavelengths, with electric-vector polarization angle (PA hereafter) parallel to the jet axis on the sky \citep{Liodakis2022-Mrk501}. The initial results were interpreted in terms of synchrotron radiation arising from a shock-accelerated electron population that becomes energy-stratified due to radiative cooling. The electrons propagate away from the shock front, radiating at optical and radio wavelengths, with PD decreasing as they encounter larger, and likely increasingly turbulent, regions of the jet flow.  

Following Mrk 501, observations of similar blazars with peaks of their synchrotron spectral energy distribution (SED) at ultraviolet or X-ray frequencies,\footnote{Mrk 501 is a high-synchrotron-peaked (HSP) source, defined as $\nu_{\rm syn}>$10$^{15}$ Hz \citep{Ajello2020}.} namely Mrk 421 \citep{DiGesu2022-Mrk421,DiGesu2023,kim2024}, PG~1553+113 \citep{Middei2023}, 1ES~0229+20 \citep{Ehlert2023}, and 1ES~1959+65 \citep{Errando2024}, and PKS~2155-304 \citep{Kouch2024} have provided further evidence supporting the shock-accelerated energy-stratified scenario. 

Since the first two Mrk~501 observations in March 2022, IXPE has obtained additional observations of the source accompanied by concurrent multi-wavelength data, including one in 2022 July \citep{lisalda2024} and three more in 2023. 
Here we present the analysis of the three 2023 observations and compared them to the three 2022 observations to provide the first long-term analysis of a blazar's X-ray polarimetric properties. Section~\ref{sec:data} describes all of our multi-wavelength polarization observations and in Section~\ref{sec:analysis} we discuss the X-ray polarimetric and spectroscopic data analysis. In Section~\ref{sec:ltv}, we 
analyze the long-term polarimetric behavior. We draw our conclusions in Section~\ref{sec:conclusions}. The uncertainties reported in this work are at the $68\%$ confidence-interval (1$\sigma$) unless stated otherwise.

\section{Data} \label{sec:data}
\subsection{IXPE Data}

IXPE targeted Mrk~501 three times in 2023, following three pointings in 2022, all with exposure times of $\sim100$ ks. (See Table~\ref{tab:obs} for details.) For these observations, 
we first apply the background screening criteria of \cite{dimarco2023ixpebg} to the publicly available pipeline-processed level 2 events to remove background events. 
We then extracted the polarization cube with the \texttt{PCUBE} algorithm using the {\sc xpbin} function of {\sc ixpeobssim},\footnote{Version 31.0.1, with instrument response function (IRF) v13; see \url{https://github.com/lucabaldini/ixpeobssim}.} with a $1\arcmin$ radius source region centered at the source position. 
%The extraction region, listed in Table~\ref{tab:obs}, varied slightly among the observations.
For each observation, a background polarization cube was also extracted from an annulus with $2\farcm5$ and $5\arcmin$ inner and outer radii with the same center as the source position. Section~\ref{sec:analysis} describes the procedures followed to generate the $I$, $Q$, and $U$ Stokes parameters from the extracted IXPE data.

For each \obsid, spectra files were extracted using {\sc xselect} with the same source and background regions. For the spectral response functions, we used the 20240125 version of the IXPE CALDB with the event weighing algorithm using an elongation normalized parameter of $\alpha^{0.75}$. We also accounted for the vignetting and aperture corrections to the auxiliary response functions using {\sc ixpecalcarf} within the HEASARC FTOOLS \citep{heasoft}. 
While the IXPE observations for pointed sources generally place the target at the optical axis of the mirror modules, small offsets on the scale of $\sim 1^\prime$ are expected due to IXPE's boom-drift correction and dithering effects. These effects were taken into account using IXPE's level 2 attitude files when deriving the vignetting and aperture corrections. 
We group the total flux density (Stokes {\it I}) spectra with a minimum of 25 counts per bin. The {\it Q} and {\it U} spectra were binned with a constant 5 channels (0.2 keV) per bin. 

IXPE also observed Mrk~501 for three times in 2022 (see Table~\ref{tab:obs}). 
We re-processed and extracted \pcube\ and spectra for the 2022 observations similar to how we processed the 2023 observations. One distinction is that for the first two IXPE observations in 2022, we used a fixed off-axis angle of $2\farcm73$ to calculate the vignetting and aperture correction to the response functions instead of using the attitude files. This is motivated by a correction in the IXPE optical axis relative to the star trackers took place on June 7th, 2022. For the remainder of the paper, we refer to the six IXPE observations in their chronological order (see Table~\ref{tab:obs}) as IXPE1 through IXPE6.

\subsection{Ancillary X-ray Data}
\label{sec:ancilxray}
To supplement the IXPE observations, we make use of the contemporary ancillary X-ray data from NuSTAR,and XMM-Newton when available; see Table~\ref{tab:obs}.
Mrk~501 is one of the targets in the list of Fermi-LAT ``sources of interest'' with extensive monitoring by the Neil Gehrels Swift Observatory X-Ray Telescope \citep{Stroh2013}. For the remaining IXPE observations without contemporary NuSTAR or XMM-Newton data, we make use of Swift/XRT data obtained during the IXPE pointing. The ancillary X-ray data are used for spectral and spectral-polarimetric analysis described in Section~\ref{sec:analysis}. 
Additionally, photometric measurements from the Swift Ultraviolet/Optical Telescope (UVOT) were also included in our analysis.
For NuSTAR and Swift XRT and UVOT data, {\sc heasoft} 6.31.1 was used for data reduction and extraction. 
For XRT, the Windowed Timing mode spectra were extracted using a circular region with a 47\arcsec\ radius, with backgrounds accounted for with time-dependent WT background files. 
For the Photon Counting mode observations, the spectra were extracted 
following the procedures described in \citet[][see their Section~2]{middei2022}. 
For NuSTAR, we extracted spectra from both focal plane modules (FPM) with a 90$^{\prime\prime}$ radius source aperture and a background annulus with 150\arcsec\ and 220\arcsec\ inner and outer radii, respectively. 
For XMM-Newton, XMMSAS version 20.0.0 was used for data reduction and spectral extraction of the EPIC-PN data. Background flares were screened and filtered using a 3$\sigma$ clipping algorithm. The source and background spectra were extracted from the timing-mode data with a source region of 20 detector rows and a source-free region between $3\leq {\rm RAWX} \leq 5$.

\subsection{Multi-wavelength observations}\label{sec:multi}

\begin{table*}
\footnotesize
\hspace{-4em}
\begin{tabular}{ccccccc}
\hline\hline
Observation & Optica-IR & Radio (GHz) & \pdo\ (R) & \pao\ & \pdra\ (86, 225.5 GHz)& \para\ (86, 225.5 GHz) \\
& (1) & (2) & (3) &(4) &(5) &(6)\\
\hline
IXPE1&  {\it V},{\bf B},{\it R,I,J} & {\it 86} & $6.6\pm0.4$                      &$110\pm5$ & $1.5\pm0.5$, - & $148\pm10$, -\\%& V, B, R, I, J, 86 \\
IXPE2 &  {\it V},{\bf B},{\it R,I,J,H,K} & N/A & $4.7\pm0.3$                      &$120\pm3$ & -, -& -, - \\%& V, B, R, I, J, H, K\\
IXPE3 &  {\it V},{\bf B},{\it R,I,J,H,K} & {\it 86, 225.5} & $2.7\pm0.5$          &$109\pm5$ & $1.5\pm0.3$, $1.3\pm0.3$ & $144\pm5$, $130\pm3$ \\%& V, B, R, I, J, H, K,86, 225.5  \\

IXPE4 & {\it V},{\bf B},{\it R,I},J & 4.85, {\it 10.45, 17,} & $6.6\pm0.9$ &$150\pm4$ & $3.2\pm0.6$, $2.9\pm0.5$ & $176\pm3$, $162\pm2$\\
&& {\it 86, 225.5} &&&&\\%& V, B, R, I, 10.45, 17, 86, 225.5\\
IXPE5 &  {\it V},{\bf B},{\it R,I,J} & 4.85, {\it 10.45, 17,}& $6.1\pm0.7$. &$125\pm3$ & -, $1.8\pm0.2$& -, $154\pm3$\\ 
&&{\it 22, 43}, 86, {\it 225.5}&&&& \\%& V, B, R, I, J, 10.45, 17, 22, 43, 86, 225.5\\
IXPE6 &  V,{\it R},I,J  & 4.85, {\it 10.45, 17, 22, 43,}& $5.9\pm1.5$.      &$108\pm6$ & -, $3.3\pm0.4$& -, $156\pm2$\\ 
&& 86, {\it 225.5} &&&&\\%& R, 10.45, 17, 22, 43, 86, 225.5  \\
\hline
\end{tabular}
\caption{   
Summary for the multi-wavelength data reported in this work. Column (1) shows the available optical to IR bands, while Column (2) shows the available radio bands in GHz. Shown in italicised fonts are bands at with both polarimetric and photometric measurements were available. Bands with only polarimetric measurements were shown in boldface. We also show the host-galaxy subtracted polarimetric measurements results for the R-band in columns (3) and (4). For radio polarimetric measurements, we show the polarimetric measurements from the higher frequency bands, 86~GHz and 225.5~GHz, in columns (5) and (6). 
See Section~\ref{sec:multi} for details. 
From columns (3) to (6), the PD values were shown in \%, and PA values were shown in degrees. 
}\label{tab:mwobs}
\end{table*}

Multi-wavelength observations contemporaneous with the three IXPE pointings in 2022 are described in detail in \cite{Liodakis2022-Mrk501} and \cite{lisalda2024}. Details of the analysis of data from the different telescopes can be found there as well. Here, we provide a short description and present the multiwavelength observations taken around the 2023 IXPE observation dates. 

Similar to the previous IXPE sessions, a number of telescopes operating across the electromagnetic spectrum provided contemporaneous data. Here we focus on the radio and optical regimes. For a discussion on the very high-energy $\gamma$-ray behavior of Mrk~501 during the first three IXPE observations, see \cite{Abe2024}. At radio wavelengths, those telescopes included the Effelsberg 100-m antenna, the IRAM 30~m telescope, the Korean VLBI Network (KVN), and the SubMillimeter Array (SMA; \citealp{Ho2004}). These facilities provide coverage from 4.85 to 225.5 GHz. Effelsberg observations at 4.85, 10.45, and 17~GHz were obtained as part of the Monitoring the Stokes $Q$, $U$, $I$, and $V$ Emission of AGN jets in Radio (QUIVER) program \citep{Myserlis2018,Krauss2003}. Observations at 86~GHz with the IRAM-30m telescope were taken as part of the Polarimetric Monitoring of AGN at Millimeter Wavelengths (POLAMI) project\footnote{\url{http://polami.iaa.es/}} \citep{ Agudo2018-II,Agudo2018, Thum2018}. The KVN observations used the Yonsei and Tamna antennas in single dish mode to provide observations at 22, 43, 86, and 129~GHz \citep{Kang2015}. Finally, 225.5~GHz observations were taken within the SMA Monitoring of AGNs with Polarization (SMAPOL) program (Myserlis et al., 2024 in preparation). All of the radio observations are shown in the appendix (Figure~\ref{fig:radio_obs}). The source exhibited a low degree of polarization ($<4\%$) for the entire monitoring period, with a roughly constant polarization angle fluctuating about an average of ${\rm PA}_{\rm Rad}=157\degr\pm3\degr$ at 225.5~GHz to  ${\rm PA}_{\rm Rad}=170\degr\pm1\degr$ at 4.85~GHz. We note the radio polarimetry measurements utilized in this work were all obtained with either single-dish observatories or radio telescope arrays in single-dish modes. The observed radio emission is thus dominated by the unresolved blazar jet emission and the fluxes and polarization measurements are not affected by the different aperture sizes between radio observatories. However, radio observations at higher frequencies tend to have less significant Faraday depolarization \citep{1966MNRAS.133...67B}, therefore we primarily consider the high frequency polarization measurements (see Table~\ref{tab:mwobs}).

\begin{figure}
%\DeclareGraphicsExtensions{png}
\includegraphics[width=0.5\textwidth]{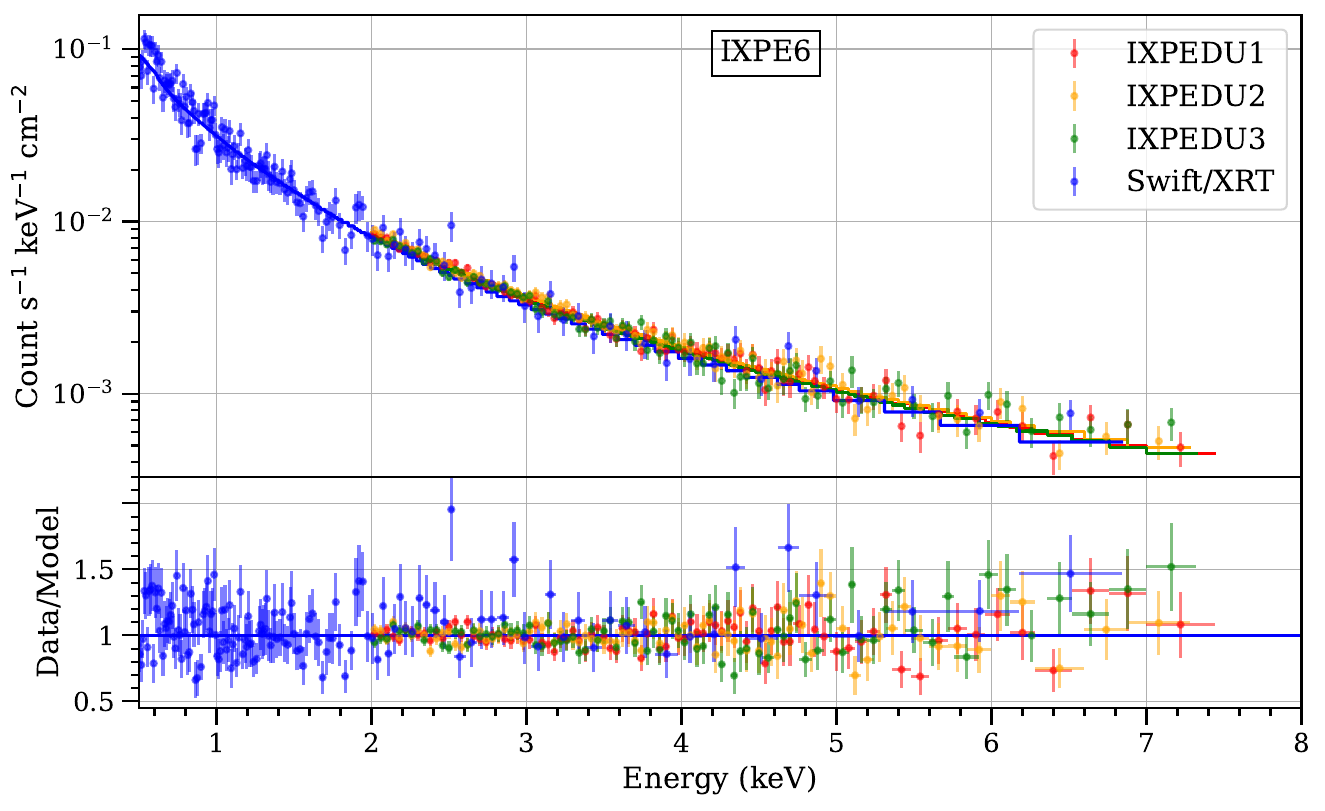}
\includegraphics[width=0.5\textwidth]{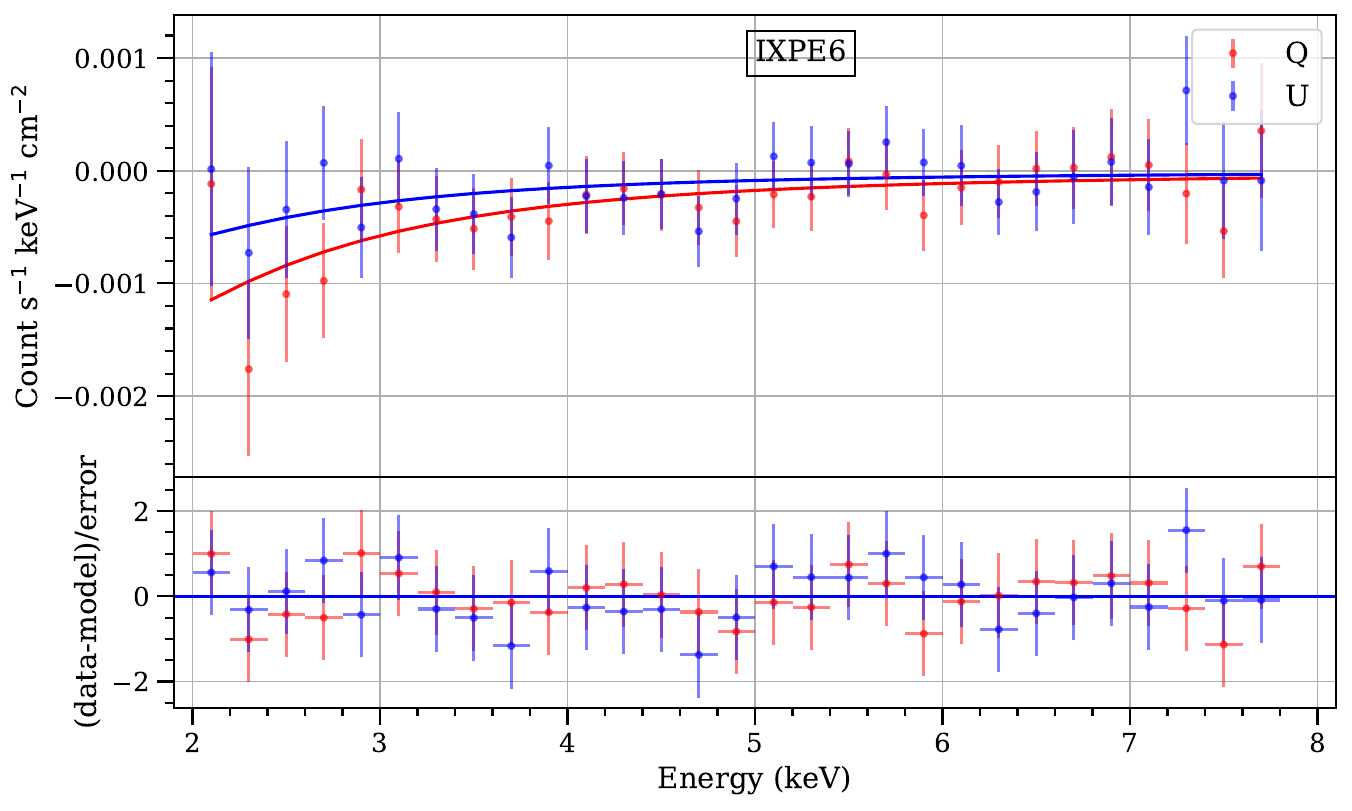}
\caption{Spectra during IXPE6 of Mrk~501 with ancillary X-ray data. The solid lines are the best-fit models. Top: Swift/XRT and IXPE total flux density spectra. Bottom: IXPE {\it Q} and {\it U} spectra for DU1.}
\label{fig:specplots}
\end{figure}

In the optical and infrared regime, Mrk 501 was observed using the Calar Alto Observatory's 2.2~m telescope, the Haleakala Observatory's T60, the KANATA telescope, LX-200 operated by St.\ Petersburg State University,  the Nordic Optical Telescope (NOT), the 1.8~m Perkins Telescope (PTO, Boston University), and the Sierra Nevada Observatory's (SNO) T90 telescope. The Calar Alto observations used the Calar Alto Faint Object Spectrograph (CAFOS) to provide R-band observations. T60 used the DiPol-2 polarimeter \citep{Piirola2014,Piirola2020}, which can simultaneously measure the PD in the B, V, R bands. LX-200 provided R-band polarimetry and B, V, R, I photometry.  HONIR (Hiroshima Optical and Near-InfraRed camera\footnote{\url{http://hasc.hiroshima-u.ac.jp/instruments/honir/filters-e.html}}) at the KANATA telescope \citep{Kawabata1999,Akitaya2014} can simultaneously measure polarization at multiple optical/infrared (IR) bands; it provided R and J band observations.  NOT used the Alhambra Faint Object Spectrograph and Camera (ALFOSC) and the Tuorla Observatory's semi-automatic data reduction pipeline \citep{Hovatta2016,Nilsson2018} for B, V, R, I observations. PTO used the PRISM camera\footnote{\url{https://www.bu.edu/prism/}} for polarimetric (R) and photometric (B, V, R, I) observations. The T90 observations at SNO were performed with a set of polarized filters in R band. Several observations were obtained within a single night, which were then binned using a weighted average. All of the optical and infrared observations are shown in Figure~\ref{fig:radio_obs}-Right.

Unpolarized flux from the host galaxy of Mrk 501 contributes significantly to the optical/IR emission. This leads to depolarization, which we correct by subtracting the host-galaxy flux within a given aperture \citep{Nilsson2007} following \cite{Hovatta2016}. Only the R-band observations are corrected, since we lack a model for the host-galaxy light for the other bands. Instrumental limitations prevent us from correcting the T60 observations. All the other R-band observations have been corrected for the host-galaxy depolarization.  

We were not able to obtain J-band observations simultaneous with the IXPE pointings; however, the two available observations between the fourth and fifth IXPE exposures appear consistent with the optical bands. The observed PD is consistent within uncertainties in all optical/IR bands, and appears to vary in tandem. The intrinsic R-band PD (${\rm PD}_{\rm opt}$ hereafter) varies from about 1.6\% to $\sim$9\%, with a median of 6.6\%$\pm$1.9\%. The PA (${\rm PA}_{\rm Opt}$ hereafter) fluctuates around the jet axis position angle on the plane of the sky \cite[$120\degr\pm12\degr$;][]{Weaver2022} from  $\sim100\degr$  to $\sim150\degr$, with a median of $133\degr\pm14\degr$.
%For the IXPE4 we find ${\rm PD}_{\rm opt}=6.1\%\pm1.9\%$ along ${\rm PA}_{\rm Opt}=139\degr\pm8\degr$, for IXPE5 ${\rm PD}_{\rm opt}=7.3\%\pm0.8\%$ along ${\rm PA}_{\rm Opt}=148\degr\pm3\degr$, while for IXPE6 ${\rm PD}_{\rm opt}=7.1\%\pm0.8\%$ along ${\rm PA}_{\rm Opt}=119\degr\pm9\degr$.  
A summary of the multiwavelength observations is provided in Table~\ref{tab:mwobs}.

\section{Analysis}\label{sec:analysis}

\begin{table*}
\begin{tabular}{ccccccc}
\hline \hline
IXPE Observation & $\chi^2$/d.o.f. & $\alpha$  & $\beta$  & norm  & PD (\%) & PA (deg)  \\
\hline  
%IXPE1 & 1285.46 / 1274 & $1.75^{+0.04}_{-0.04}$ & $0.36^{+0.02}_{-0.02}$ & $0.043^{+0.001}_{-0.001}$ & $9.79^{+1.66}_{-1.65}$ & $-43.74^{+4.86}_{-4.86}$ \\
%IXPE2 & 1654.83 / 1480 & $1.5^{+0.03}_{-0.03}$ & $0.4^{+0.02}_{-0.02}$ & $0.066^{+0.001}_{-0.001}$ & $10.3^{+1.38}_{-1.38}$ & $-65.14^{+3.85}_{-3.84}$ \\
%IXPE3 & 559.2 / 643 & $1.99^{+0.02}_{-0.02}$ & $0.36^{+0.03}_{-0.03}$ & $0.053^{+0.001}_{-0.001}$ & $6.86^{+1.83}_{-1.83}$ & $-46.33^{+7.72}_{-7.72}$ \\
%IXPE4 & 582.1 / 577 & $2.28^{+0.0}_{-0.0}$ & $0.1^{+0.01}_{-0.01}$ & $0.037^{+0.0}_{-0.0}$ & $8.96^{+2.41}_{-2.41}$ & $-70.29^{+7.82}_{-7.8}$ \\
%IXPE5 & 594.88 / 589 & $2.27^{+0.01}_{-0.01}$ & $0.03^{+0.01}_{-0.01}$ & $0.041^{+0.0}_{-0.0}$ & $5.98^{+2.13}_{-2.13}$ & $-72.67^{+10.5}_{-10.44}$ \\
%IXPE6 & 603.62 / 599 & $2.15^{+0.02}_{-0.02}$ & $0.2^{+0.03}_{-0.03}$ & $0.041^{+0.001}_{-0.001}$ & $18.48^{+2.19}_{-2.18}$ & $-76.84^{+3.38}_{-3.38}$ \\
%IXPE1 & 1285.46 / 1274 & $1.75\pm0.04$ & $0.36\pm0.02$ & $0.043\pm0.001$ & $9.8\pm1.7$ & $-44\pm5$ \\
%IXPE2 & 1654.83 / 1480 & $1.50\pm0.03$ & $0.40\pm0.02$ & $0.066\pm0.001$ & $10.3\pm1.4$ & $-65\pm4$ \\
%IXPE3 & 559.2 / 643 & $1.99\pm0.02$ & $0.36\pm0.03$ & $0.053\pm0.001$ & $6.9\pm1.8$ & $-46\pm8$ \\
%IXPE4 & 582.1 / 577 & $2.28\pm0.01$ & $0.10\pm0.01$ & $0.037\pm0.001$ & $9.0\pm2.4$ & $-70\pm8$ \\
%IXPE5 & 594.88 / 589 & $2.27\pm0.01$ & $0.03\pm0.01$ & $0.041\pm0.001$ & $6.0\pm2.1$ & $-73\pm11$ \\
%IXPE6 & 603.62 / 599 & $2.15\pm0.02$ & $0.20\pm0.03$ & $0.041\pm0.001$ & $18.4\pm2.2$ & $-77\pm3$ \\
IXPE1 & 1285.46 / 1274 & $1.75\pm 0.04$ & $0.36\pm0.02$ & $0.043\pm0.001$ & $9.8\pm1.7$ & $-44\pm5$ \\
IXPE2 & 1654.83 / 1480 & $1.50\pm 0.03$ & $0.40\pm0.02$ & $0.066\pm0.001$ & $10.3\pm1.4$ & $-65\pm4$ \\
IXPE3 & 559.2 / 643 & $1.99\pm 0.02$ & $0.36\pm0.03$ & $0.053\pm0.001$ & $6.9\pm1.8$ & $-46\pm8$ \\
IXPE4 & 582.1 / 577 & $2.28\pm 0.00$ & $0.10\pm0.01$ & $0.037\pm0.000$ & $9.0\pm2.4$ & $-70\pm8$ \\
IXPE5 & 594.88 / 589 & $2.27\pm 0.01$ & $0.03\pm0.01$ & $0.041\pm0.000$ & $6.0\pm2.1$ & $-73\pm11$ \\
IXPE6 & 603.62 / 599 & $2.15\pm 0.02$ & $0.2\pm0.03$ & $0.041\pm0.001$ & $18.5\pm2.2$ & $-77\pm3$ \\
\hline
\end{tabular}
\caption{Spectral-polarimetric results with IXPE and ancillary X-ray data for each ObsID, including the fitting statistics and degrees of freedom (d.o.f.), log-parabolic model parameters $\alpha$, $\beta$, and normalization (norm), and the polarization degree and angle (PD and PA). The uncertainties quoted here are for 68\% confidence intervals.}
\label{tab:xspeca}

\begin{tabular}{ccccccc}
%Combined & 9 $\pm$ 1 & $-61.31\pm$ 2.34 & 1406.53/1341 & 1.92 $\pm$ 0.03 & 0.38 $\pm$ 0.07 & 0.011    \\
\hline \hline
 & \multicolumn{2}{c} {PCUBE} & \multicolumn{2}{c} {Spectro-polarimetry (all)} & \multicolumn{2}{c} {Spectro-polarimetry (IXPE-only)}\\
\cline{2-7}
 & PD (\%) & PA (deg) & PD (\%) & PA (deg)& PD (\%) & PA (deg)\\
\hline
IXPE1 & $10.1\pm2.1$ & $-53\pm6$& 
$9.8\pm1.7$ & $-44\pm5$  &
$9.0\pm2.6$&$-44\pm8$\\
IXPE2 & $11.4\pm1.7$ & $-64\pm4$ & 
$10.3\pm1.4$ & $-65\pm4$  &
$9.4\pm2.0$&$-64\pm6$\\
IXPE3 & $5.9\pm2.3$ & $-40\pm11$ &
$6.9\pm1.8$ & $-46\pm8$  &
$6.9\pm2.9$&$-48\pm12$\\
IXPE4 & $9.9\pm3.2$& $-75\pm9$ &
$9.0\pm2.4$ & $-70\pm8$ &
$9\pm4$&$-71\pm16$\\
IXPE5 &$6.7\pm2.8$& $-84\pm12$ &
$6.0\pm2.1$ & $-73\pm11$ &
$5.9\pm3.4$&$-72\pm17$\\
IXPE6 & $15.3\pm2.8$& $-82\pm5$ &
$18.5\pm2.2$ & $-77\pm3$ &
$16.3\pm3.4$&$-76\pm6$\\
Combined & $8.9\pm1.0$ & $-65\pm3$ &
 - & - &
$8.7\pm1.2$ & $-62\pm4 $
\\
\hline
\end{tabular}
\caption{Comparison between polarization measurements with \texttt{PCUBE} analysis, and spectral-polarimetric analysis with and without ancillary X-ray data.}
\label{tab:xspecb}
\end{table*}

\begin{figure}
\includegraphics[width=\columnwidth]{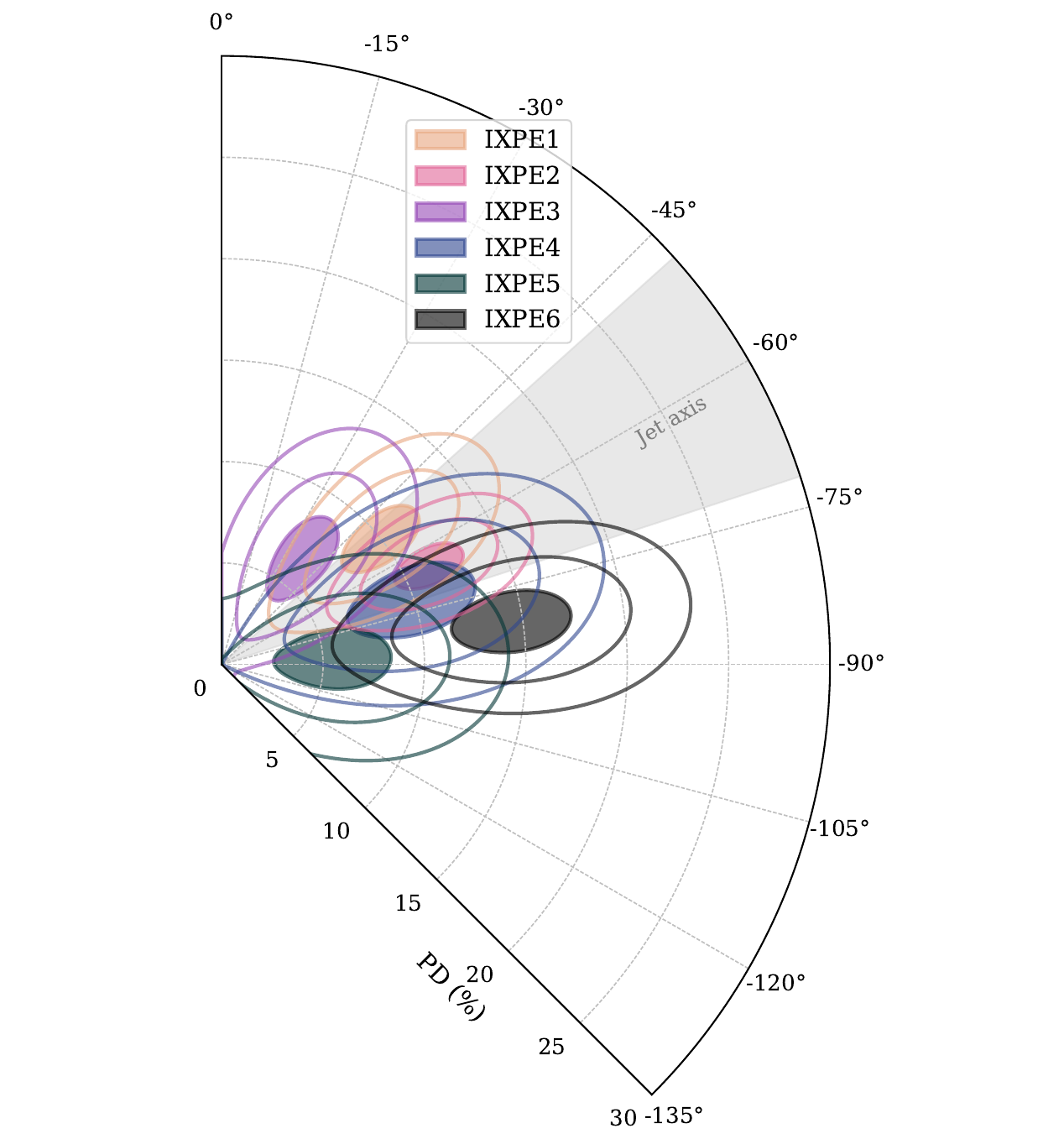}
\includegraphics[width=\columnwidth]{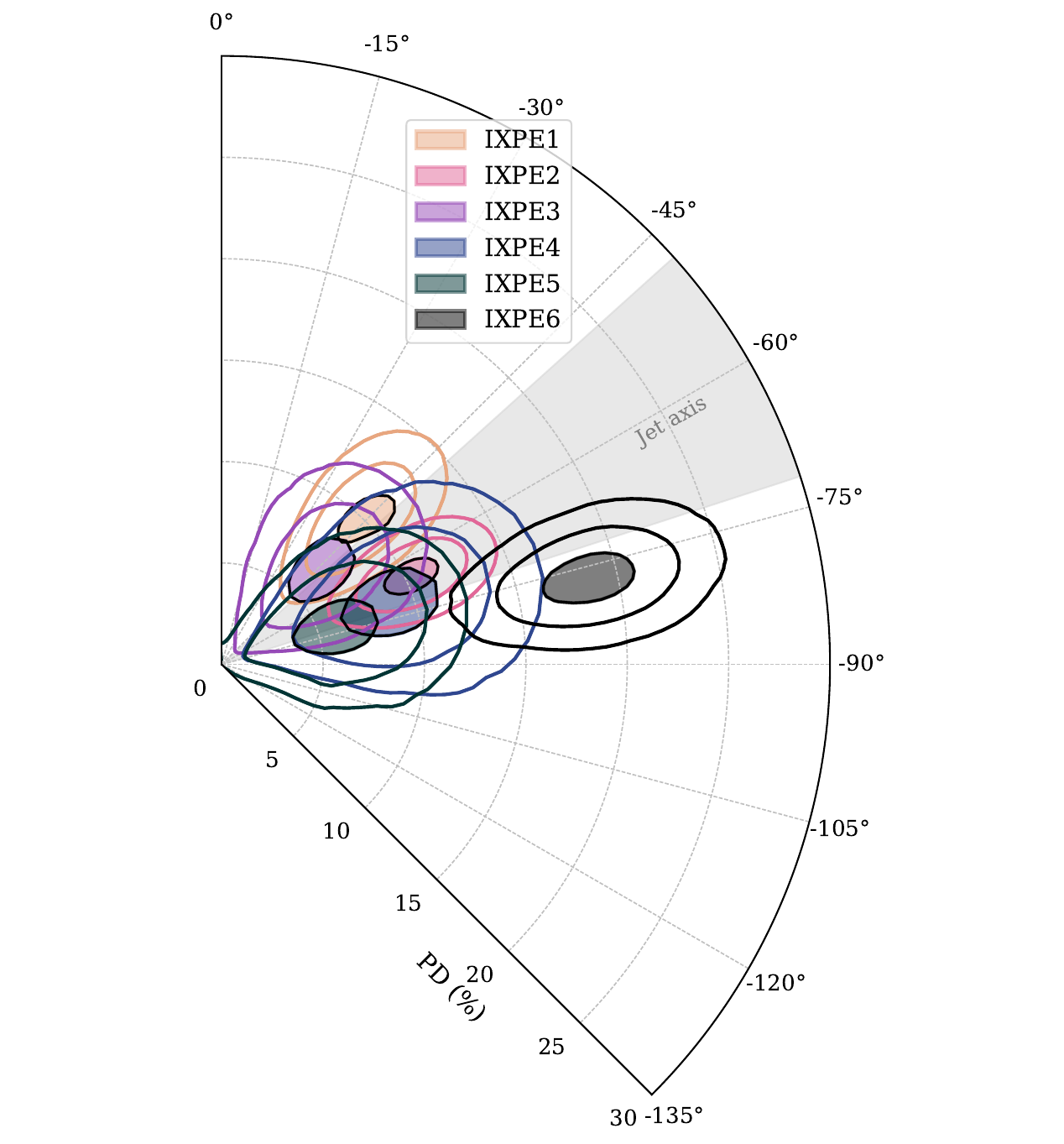}
\caption{
Polar plots displaying polarization measurements based on the model-independent \texttt{PCUBE} analysis (top). The contours represent the 1$\sigma$, 2$\sigma$, and 3$\sigma$ uncertainties in ${\rm PD}_{\rm X}$ and ${\rm PA}_{\rm X}$ based on the $Q,U$ error circles and their relations to ${\rm PD}_{\rm X}$ and ${\rm PA}_{\rm X}$ described in Section~\ref{sec:pcube}. 
The spectro-polarimetric fitting results with IXPE and ancillary X-ray data analysis are shown in the bottom. The error contours are based on the MCMC sampling described in Section~\ref{sec:specP} within 68\%, 95\%, and 99.7\% ranges.  
The results obtained from these two independent methods are consistent with each other.}
\label{fig:polarplot}
\end{figure}

\begin{figure}
\DeclareGraphicsExtensions{png}
\includegraphics[width=0.42\textwidth]{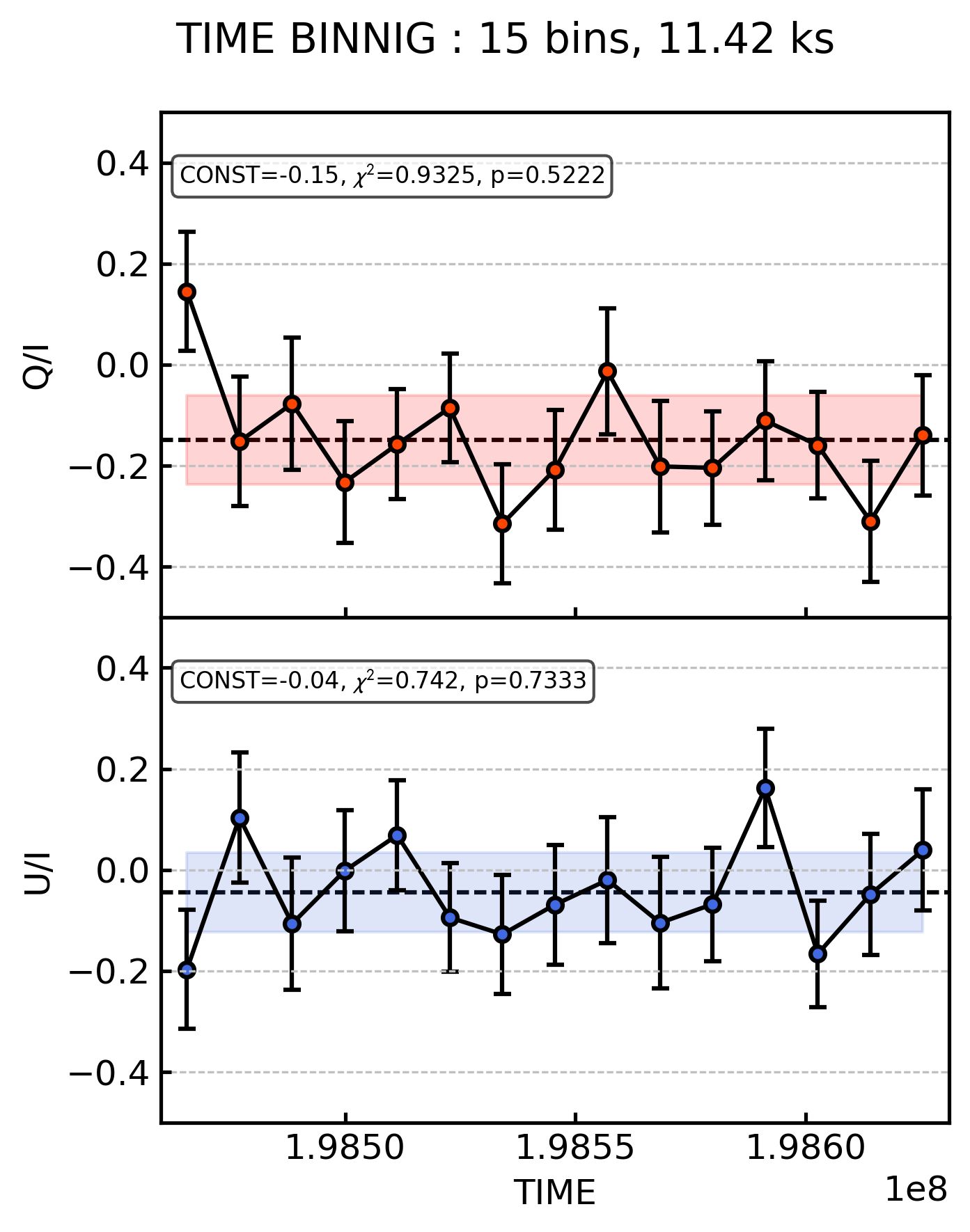}
\includegraphics[width=0.52\textwidth]{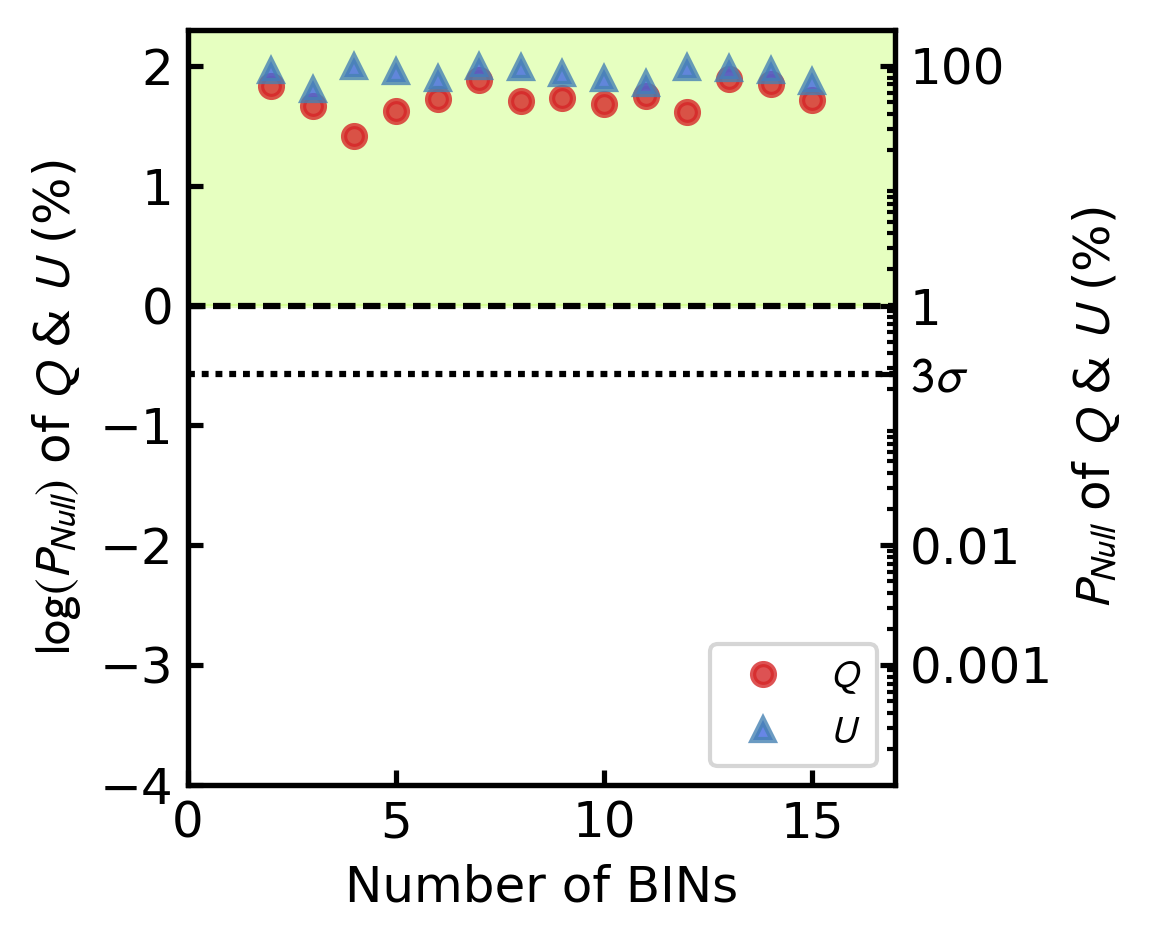}
\caption{
Top: light curves for the Stokes parameters {\it Q} and {\it U} for IXPE6; the dotted lines and the shaded regions represent the time-averaged values. Bottom: 
null-hypothesis probability ($P_{\rm Null}$) of the $\chi^2$ test for the variability of {\it Q} and {\it U} for IXPE6. 
The green shaded area above the dashed line makrs the region with $P_{\rm Null}>1\%$. 
The dotted line represents the 3$\sigma$ (99.73\%) significance level. 
For the range of bin sizes explored here, we find no statistical evidence ($<3\sigma$) that Mrk~501 varied in either  {\it Q} or  {\it U}. Similar results were found for all six Mrk~501 IXPE observations discussed in this work. 
}
\label{fig:singlevar}
\end{figure}

\subsection{Spectro-polarimetry}\label{sec:specP}
For the spectro-polarimetric analysis, we combine a log-parabola spectrum \citep{mass04logpar} with a multiplicative model that applies a constant polarization, {\sc polconst}. 
The model used in {\sc xspec} can be written as follows: 
\[
{const\times TBabs\times polconst\times logpar}.
\]
The {\sc const} component accounts for the instrumental normalization. The {\sc polconst} component includes two parameters: the polarization degree ${\rm PD}_{\rm X}$ and the polarization angle ${\rm PA}_{\rm X}$. 
The log-parabolic model is defined as the 
$F(E) = K(E/E_{\rm pivot})^{-(\alpha+\beta\log(E/E_{\rm pivot}))}$. The parameters $\alpha$, $\beta$, and $E_{\rm pivot}$ defines the energy-dependent power-law index of the spectrum, and $K$ represents the normalization. $E_{\rm pivot}$ is a constant for optimal fitting results in the log space, here we fixed to 1~keV.  
The results of the analysis, including both IXPE and ancillary soft X-ray data, are listed in Table~\ref{tab:obs}. The XMM-Newton and Swift/XRT data were limited to 0.5--10 keV for the model fitting. For NuSTAR, the 3--30~keV data were used (higher energy channels were not used due to the limited signal to noise ratio). 
For spectral fitting, we used an MCMC sampling approach with the Goodman-Weare algorithm and 500,000 steps.
As a demonstration, we plot the spectra and the data/model ratio for IXPE6 in Figure~\ref{fig:specplots}. The {\it I, Q, and U} spectral plots for the other IXPE observations are shown in the Appendix in Fig.~\ref{fig:appspecplots} and Fig.~\ref{fig:appquplots}. We also repeated the analysis using only the IXPE data, and found the results to be generally consistent with the analysis with the ancillary X-ray data.

The fitting parameters are listed in Table~\ref{tab:xspeca}. For comparison, the polarization degree and angle from the \texttt{PCUBE} analysis and the spectro-polarimetric analysis are listed in Table~\ref{tab:xspecb} and shown in Fig.~\ref{fig:polarplot}.
We also extracted time-averaged IXPE spectra by merging event lists from all six observations. Source and background spectra were generated from merging events in the source and background regions for each observation, respectively. 
The ancillary response function was calculated with the {\sc ixpecalcarf} script in a similar manner as for each observation. 

%[Consider adding a energy-dependent MDP99 plot.]

\subsection{Model-independent Polarimetry with IXPE}\label{sec:pcube}
Here we describe the results from the polarization cube analysis, 
which calculates the X-ray PD (\pdx) and PA (\pax) using Stokes parameters $I$, $Q$, and $U$ for all of the events within the source extraction region, where ${\rm PD}_{\rm X}=\sqrt{(Q/I)^2 + (U/I)^2}$ and ${\rm PA}_{\rm X}=0.5\arctan(U/Q)$ \citep{kislat2015}.
Background subtraction was accomplished by subtracting the sums of the Stokes parameters in the background region from those in the source region. We report the model-independent polarization measurement events in the 2--8 keV range in Table~\ref{tab:xspecb}. The minimum detectable polarization at 99\% confidence interval (${\rm MDP}_{99}$), defined as ${\rm MDP}_{99}\approx 4.29/(\sqrt{C_s}\langle\mu\rangle$), is dependent on the mean count-weighted modulation factor $\langle\mu\rangle$ and the total source photon counts $C_s$. The 2--8~keV \mdp\ is listed in Table~\ref{tab:obs}. 
To test the feasibility of energy-resolved polarimetry, we divided IXPE6 into two energy bins covering the range 2--4~keV and 4--8~keV. IXPE6 has the highest 2--8~keV PD among all IXPE observations at $15.3\%\pm2.8\%$, with ${\rm MDP}_{99}=8.5\%$. However, \mdp\ over 4--8~keV increases to $21\%$, and we did not find significant polarization detection in this high-energy bin. Similar results with 4--8~keV PD $< {\rm MDP}_{99}$  were found for the other IXPE observations. Therefore, we do not include energy-resolved analysis in this work. 

In summary, we detected polarization at $>3\sigma$ significance 
for the first two and the last IXPE observations both with the spectro-polarimetric or \texttt{PCUBE} analysis. The detection significance of the other three observations was slightly less than $3\sigma$. 

\subsection{Polarization Variability}

For individual IXPE observations, we investigated whether the polarization varied on various time scales within each observation. This is motivated by the recent observations of another HSP blazar, Mrk~421, which showed a rotation in the polarization angle during one of its observations \citep{DiGesu2023}. 
The details of the timing analysis are described in Section~2.2 of \cite{kim2024}. In short, for each observation, the data were divided into $N$ bins with identical widths in time, where $N$ ranged from 2 to 15. For a 100~ks observation, this means that the time in each bin ranges from 50~ks to 6.7~ks. For each binning scheme, the normalized Stokes parameters {\it Q} and {\it U} were calculated. The distributions of {\it Q} and {\it U} as a function of time were then compared to the assumption that {\it Q} and {\it U} are constant over time. See Figure~\ref{fig:singlevar} for an example. For each bin, the comparison was done with a $\chi^2$ test. We found that all six observations have {\it Q} and {\it U} distributions consistent with the time-independent assumption (i.e., the null hypothesis $P_{\rm Null} > 1\%$; see Figure~\ref{fig:singlevar}), suggesting that the 2--8 keV polarization of Mrk~501 did not vary significantly during each of the IXPE observations reported here. 
%We did not investigate energy-dependent PD within individual IXPE observations, given the fact that even in the full band the detection significance of X-ray PD is already close to \mdp.

\section{Long-term variability}\label{sec:ltv}
\begin{figure}
\includegraphics[width=\columnwidth]{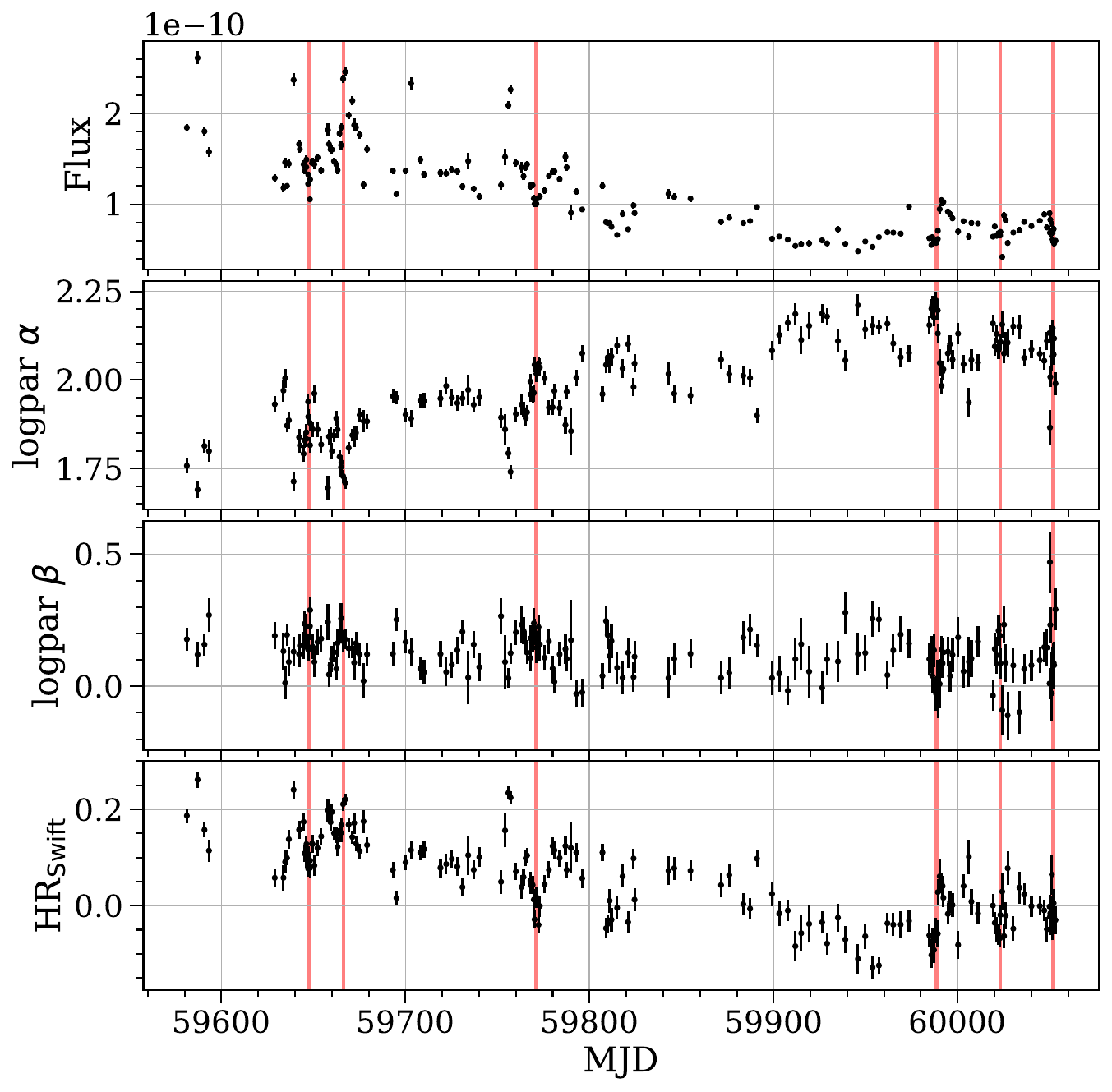}
\caption{Swift/XRT X-ray spectral properties as a function of time in MJD. From top to bottom, the panels include the 2--8~keV flux in $10^{-10}$ \fluxcgs, the log-parabolic model paramters $\alpha$ and $\beta$, and the spectral hardness ratio defined as (H$-$S)/(H+S), where H and S are Swift fluxes in the 2--10~keV and 0.5--2~keV bands, respectively. Dates of IXPE observations are highlighted in red. 
}
\label{fig:ltswift}
\end{figure}

\begin{figure}
\includegraphics[width=\columnwidth]{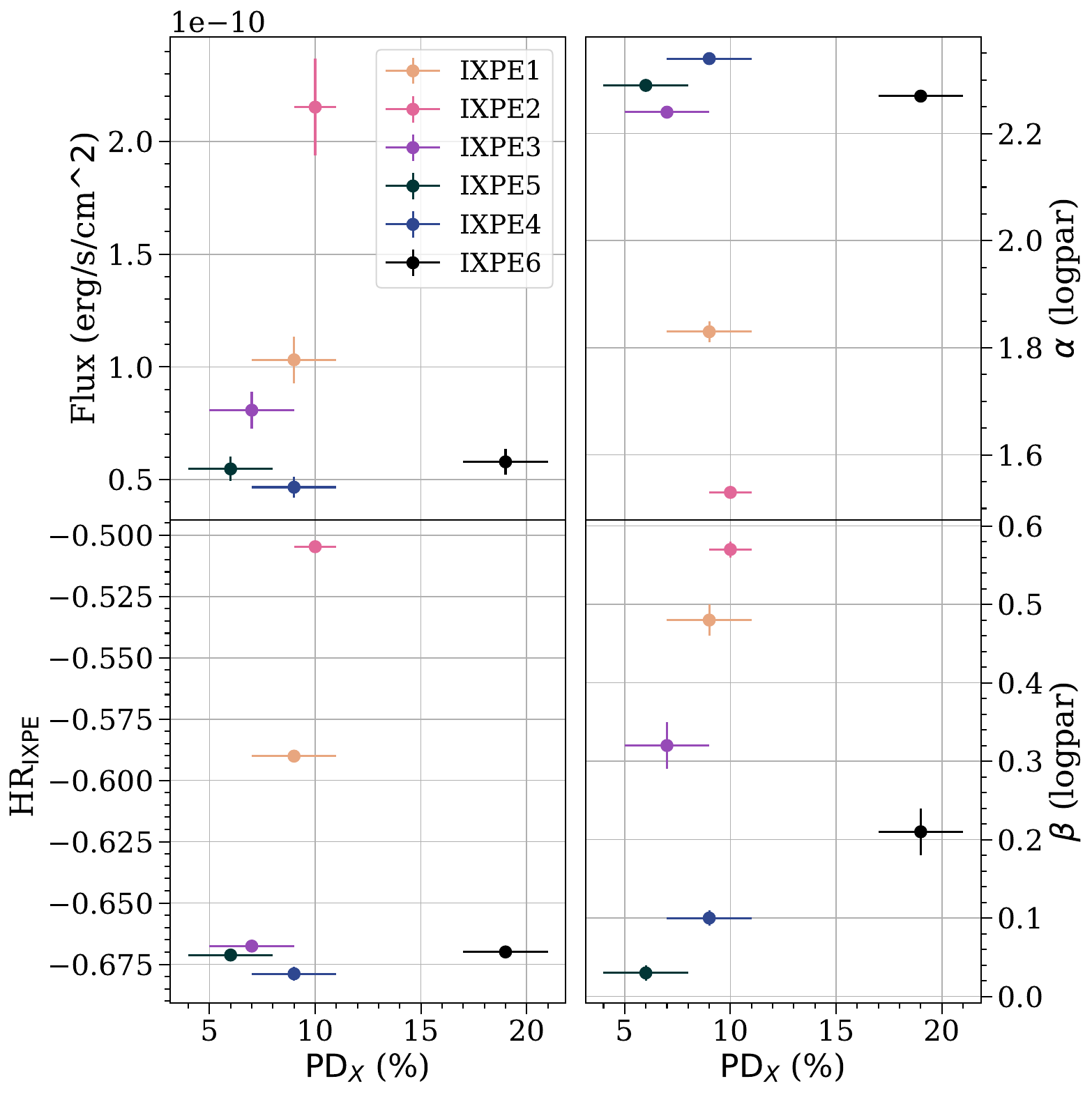}
\caption{Various X-ray spectral properties as a function of the X-ray PD, including the IXPE 2--8~keV flux (top-left), photon index $\alpha$ (top-right), IXPE 2--4 and 4--8~keV hardness ratio (${\rm HR}_{\rm IXPE}$, bottom-left, see Section~\ref{sec:xpol} for definitions), and spectral curvature (bottom-right).}
\label{fig:pdvsall}
\end{figure}

\subsection{X-ray}\label{sec:xpol}

We first show the spectral properties computed from the 167 Swift/XRT observations described in Section~\ref{sec:data} and plotted in Figure~\ref{fig:ltswift}. 
The fluxes were obtained by fitting the 0.3--10~keV Swift/XRT data with a log-parabola model with a pivotal energy set to 1~keV. Photo-electric absorption from Galactic gas with a column density $N_{\rm H}=1.69\times10^{20}$~cm$^{2}$ was included in the spectral fits. During the time between IXPE1 and IXPE6, the Swift/XRT X-ray fluxes ranged between $ -10.38 < \log F_{\rm 0.3-10 keV} < -9.58$ (\fluxcgs), with the flux peaking at the first two IXPE observations, then gradually declining into a relatively quiescent state seen in the last three observations. The log-parabola spectral parameters, $\alpha$ and $\beta$, and the hardness ratio (${\rm HR}_{\rm XRT}$ as (H$-$S)/(H+S), where H and S are Swift/XRT 2--10 keV and 0.5--2~keV fluxes, respectively) are also shown in Figure~\ref{fig:ltswift}. Generally, Mrk~501 followed the typical ``harder-when-brighter'' behavior \citep[e.g.,][]{2005ApJ...629..686Z}. 
We also show the relation between the X-ray PD and IXPE 2--8~keV flux, hardness ratio (${\rm HR}_{\rm IXPE}$, defined as (H$-$S)/(H+S), where S and H are IXPE fluxes between 2--4 and 4--8~keV, respectively),  $\alpha$ and $\beta$ in Figure~\ref{fig:pdvsall}. 

The first three IXPE observations were discussed in \cite{Liodakis2022-Mrk501} and \cite{lisalda2024}, who found Mrk~501 to have X-ray polarization angles roughly parallel to the radio jet axis. The X-ray emission was polarized at 
${\rm PD}_{\rm X}\sim 10\%$ in the first two IXPE observations \citep{Liodakis2022-Mrk501}. The third IXPE observation formally had a PD that was slightly smaller but still consistent with the first two observations within the uncertainties.
During 2023, Mrk~501 entered a lower flux state (Figure~\ref{fig:ltswift}). The lower number of counts, softer spectra, and corresponding \mdp\ resulted in only marginal detection of polarization during IXPE4 and IXPE5, while the position angle of polarization remained parallel to the jet axis within uncertainties. 
Intriguingly, the last 2023 observation (IXPE6) showed the highest ${\rm PD}_{\rm X}$ since the launch of IXPE, while maintaining a similar PA within the uncertainties. The spectro-polarimetric analyses suggest that IXPE6 is outside of the 3$\sigma$ uncertainty range compared with IXPE1 to IXPE3.
To quantify the deviation of IXPE6 from the other 5 IXPE observations, we compare the PA and PD measurements of IXPE6 with the average of the other 5 observations with a $\chi^2$ test. This is based on the assumption that the other 5 observations have similar PA and PD distributions (see Figure~\ref{fig:polarplot}).
The test has a result of $\chi^2=17.08$. In the two degree-of-freedom case, this corresponds to a p-value of 0.0002 ($>3\sigma$), rejecting the null hypothesis that the elevated PD for IXPE6 is simply due to statistical fluctuations. 

We also explored whether the Stokes parameters of the last IXPE observation have a different distribution than the rest of the observations via two-sample, two-dimensional Kolmogorov–Smirnov (K-S) tests \citep{peacock1983}. We calculated the K-S test between the $Q$ and $U$ parameters of each two pairs of IXPE observations. Most of the combinations have a very small p-value $<0.05$, rejecting the null hypothesis that their Stokes parameters have different distributions. This is not surprising, as the non-parametric statistics we tested would be dominated by the unpolarized events that should have insignificant differences between observations due to the high X-ray PD. 
Despite the substantial increase in PD for IXPE6, we find no strong correlation between PD and other spectral properties of IXPE, as demonstrated in Figure~\ref{fig:pdvsall}.

%{\color{red} None of the Mrk 501 observations shows any evidence for either short-term (within the IXPE exposure) variability or of energy dependence in the polarization degree within the 2-8 keV band.}

\begin{figure*}
\includegraphics[width=\textwidth]{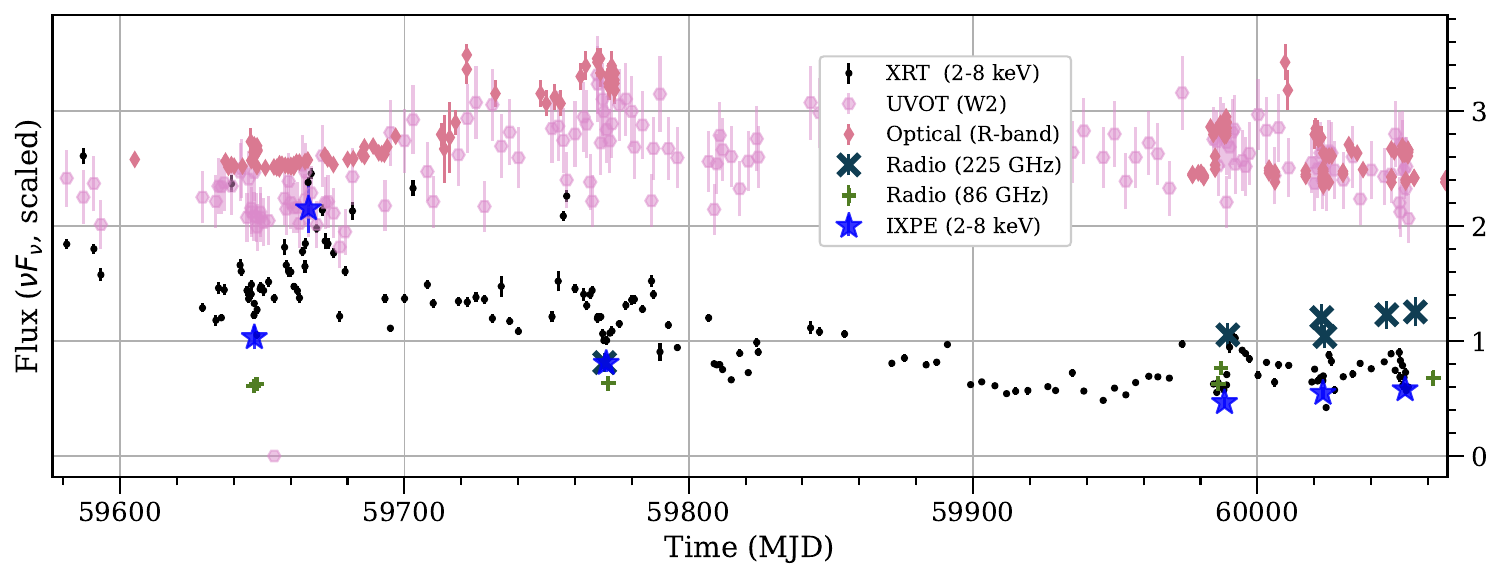}
\caption{Multiwavelength Mrk~501 variability of fluxes spanning the six IXPE observations. 
The fluxes in each band are normalized by $10^{-10}$, $10^{-11}$, $10^{-11}$, and $10^{-12}$ erg~s$^{-1}$~cm$^2$ for X-ray, UV, optical R, and radio bands, respectively.
}
\label{fig:fluxlt}
\end{figure*}

\subsection{Multiwavelength Properties}

\begin{figure}
\includegraphics[width=0.5\textwidth]{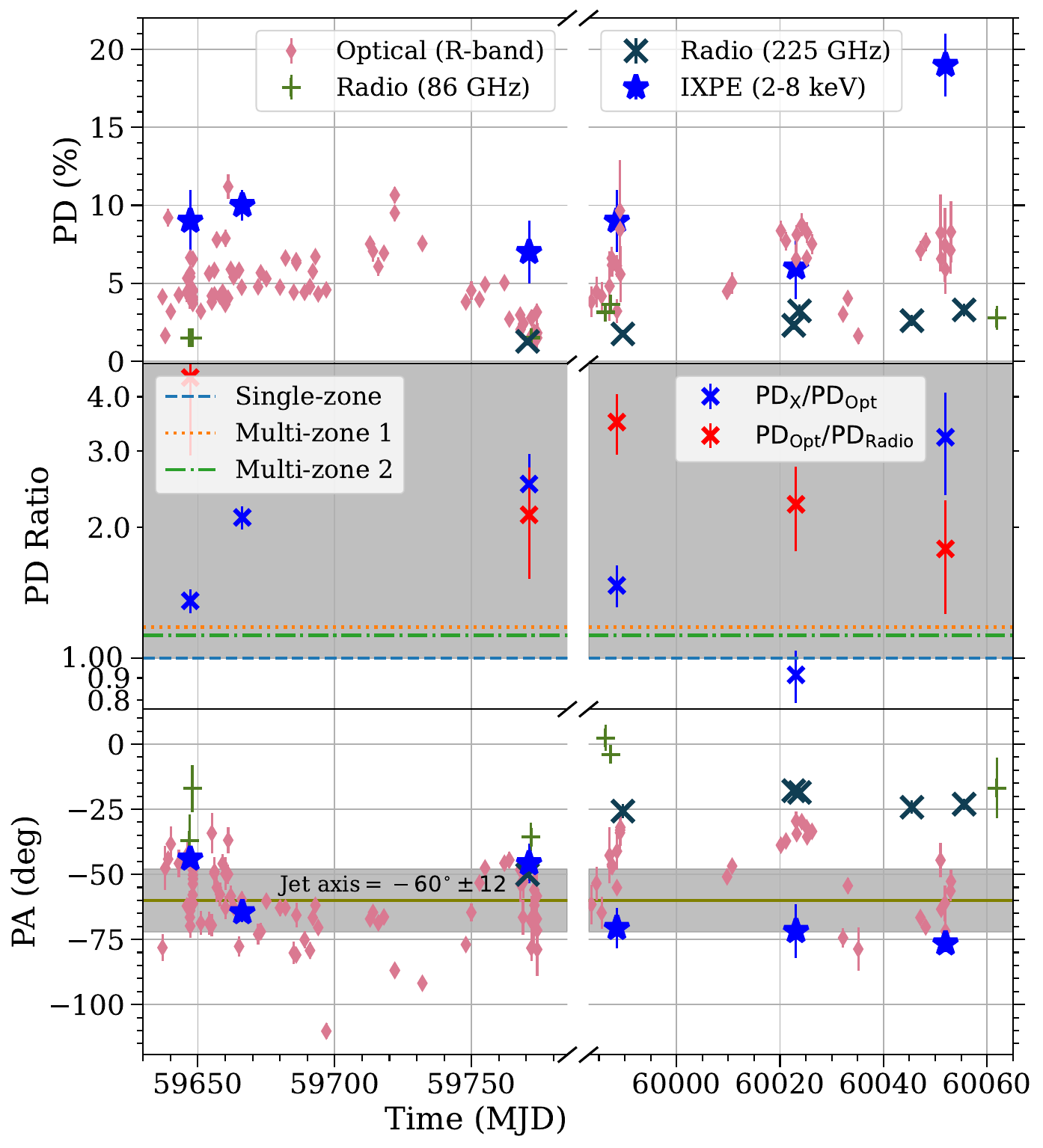}
\caption{Multiwavelength variability of Mrk~501: polarization degree PD (top), PD ratios between different bands (middle), and polarization angle PA (bottom), spanning the six IXPE observations. 
For the middle panel, the PD ratios shown here are the X-ray to optical R-band PD ratios (\pdx/\pdo), and optical R-band to radio 225 GHz (except for IXPE1 where only the 86 GHz PD measurement was available) PD ratios (\pdo/\pdra).
For the \pdx/\pdo in the middle panel, we also marked the regions 
expected based on energy-stratified shocks as grey. Expected values from a single-zone jet model \citep{2022A&A...662A..83D}, and two turbulent multi-zone jet models, including \citet[][Multi-zone 1]{2014ApJ...780...87M} and \citet[][Multi-zone 2]{2019ApJ...885...76P}, were also shown.}
\label{fig:ltpol}
\end{figure}

We plot the multiwavelength flux variability spanning the time between the IXPE1 and IXPE6 observations in Figure~\ref{fig:fluxlt}, including X-ray fluxes in the IXPE band pass (2--8~keV) from 167 Swift/XRT exposures taken between 2022-01-01 and 2023-04-15.  IXPE fluxes derived from the best-fit models are also shown in the same plot, as are the fluxes in the optical R band and radio 86 GHz and 225 GHz bands. The R-band observations were host-galaxy subtracted.
We also show the polarization properties measured by IXPE, R-band, 225~GHz, and 86~GHz in Figure~\ref{fig:ltpol}, including the time-dependence of PD, X-ray to optical PD ratio, and PA. 

\begin{figure*}
\DeclareGraphicsExtensions{png}
\includegraphics[width=\textwidth]{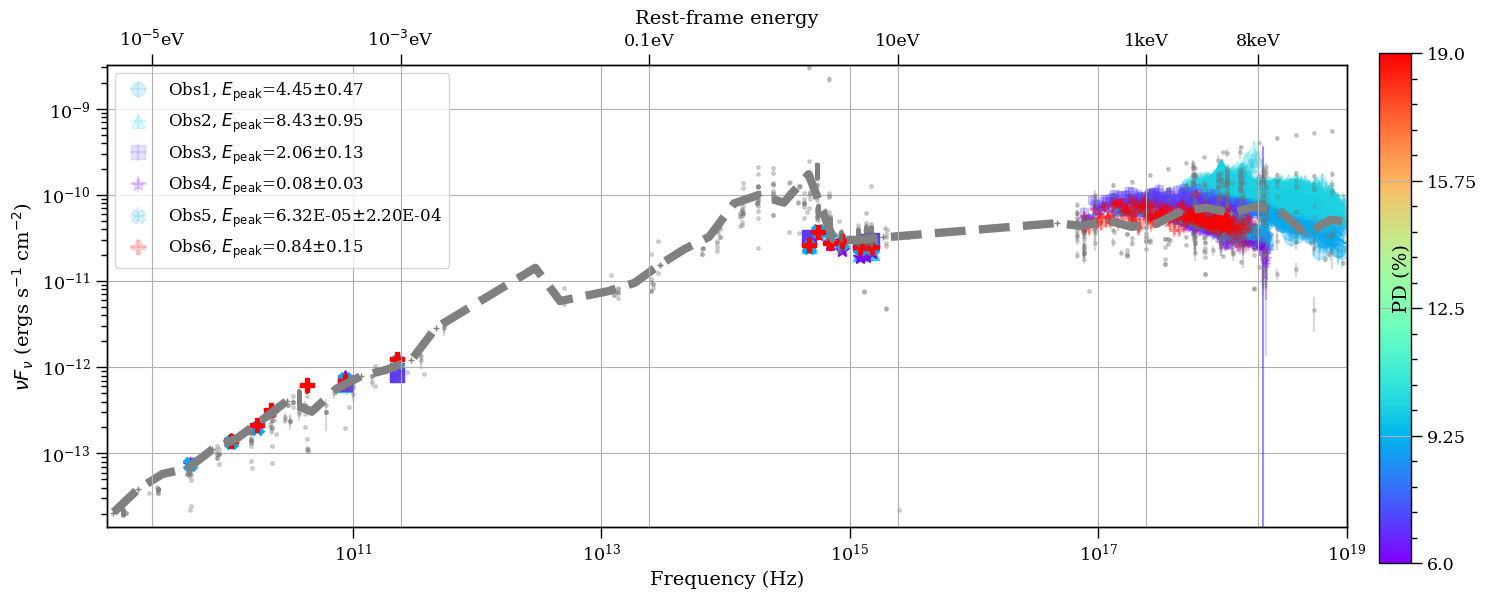}
\caption{Broad-band SED of Mrk~501, with best-fit log-parabolic model from X-ray spectro-polarimetric fitting, superposed on archival data (gray points) from the SSDC SED Builder \citep[][\url{https://tools.ssdc.asi.it/SED/}]{sedtool}. For comparison, we also display the archival data binned by 0.2dex in log frequency as a dashed gray line. The vertical color bar represents the polarization degree of each observation.
}\label{fig:broadsed}
\end{figure*}

In Figure~\ref{fig:broadsed}, we display the broad-band SED of Mrk~501 during all six IXPE sessions, including the X-ray spectra from IXPE and ancillary data sets, as well as Swift/UVOT, optical, and radio photometry. For each observation, we have also calculated the peak energy of the best-fit log-parabolic model \citep{mass04logpar} used in the spectral-polarimetry analysis, with $E_{\rm peak} = E_{\rm pivot}\, 10^{(2-\alpha)/(2\beta)}$. 

In general, we find the IXPE fluxes to be within the average fluctuations typically observed from this target \cite[e.g., see][]{Liodakis2022-Mrk501}. This would suggest that all of the IXPE  observations of Mrk 501 were taken in average/quiescent state. 
Similarly, neither the radio nor optical observations show any substantial deviation from the long-term average behavior of the source (see Figure~\ref{fig:broadsed}).
For the last IXPE observation (IXPE6), the ${\rm PD}_{\rm X}$ value is elevated above 3$\sigma$ of the average value of the other five IXPE observations. Other than the elevated ${\rm PD}_{\rm X}$, we do not find any substantial differences for IXPE6 from the other observations, including the multiwavelength polarization degree and angles, X-ray spectral shapes, and multiwavelength fluxes. 

We find no apparent correlations in the time dependence of the flux, PD, or PA values at different wavelengths as shown in Figures~\ref{fig:fluxlt} and \ref{fig:ltpol}, which is in broad agreement with the energy stratified shock scenario discussed in \cite{Liodakis2022-Mrk501}, wherein the polarized emission in the optical and radio bands originate from regions larger than the region emitting polarized X-ray emission.
We also find the X-ray PD (${\rm PD}_{\rm X}$) to be higher, or (at one epoch) similar to, the R-band PD (${\rm PD}_{\rm opt}$), 
while the R-band PD is higher than the radio PD. This is again consistent with the energy-stratified shock scenario. 
For IXPE5, ${\rm PD}_{\rm opt}$ appears to be marginally higher than ${\rm PD}_{\rm X}$
%(spectro-polarimetric),
with ${\rm PD}_{\rm X}/{\rm PD}_{\rm opt} = 0.92\pm0.12$ (1$\sigma$ uncertainty). 
%{\red Insert discussion about the lack of spectral curvature for IXPE5 here. Turbulence model? erratic rapid variability for X-ray and optical PD, so generally XPD > OPD higher but not always true}
Note that IXPE5 has the second-lowest flux and the softest X-ray spectrum (see Figure~\ref{fig:pdvsall}). Therefore, given the uncertainty of the X-ray PD, IXPE5 is still statistically consistent with ${\rm PD}_{\rm X}/{\rm PD}_{\rm opt}\geq 1$.
%we cannot rule out IXPE5 being statistically consistent with $\Pi_{\rm }/\Pi_{\rm opt}>1$. 

%In summary, we detected polarization at $>3\sigma$ significance for the first two and the last IXPE observations either with the spectro-polarimetric or \texttt{PCUBE} analysis. The detection significance of the other three observations was slightly less than $3\sigma$. 

Spanning the six IXPE observations, the X-ray PD values are generally higher than ${\rm PD}_{\rm opt}$, and ${\rm PD}_{\rm opt}$ is higher than ${\rm PD}_{\rm rad}$. This is consistent with the energy-stratified shock scenario, in which the location of the X-ray emission region is closer to the shock front than that of the longer-wavelength emission.
%
%Comment by A. Marscher: The next sentence, which I have edited out, is incorrect. A plausible cause of the lower SED peak is a dectrease in the magnetic field, which causes a lower peak flux and lower peak frequency. The energy loses are then weaker, so that the thickness of the emission region is larger, not smaller.
%
%However, in this scenario, synchrotron photons with energies higher than the synchrotron peak should originate from a thin layer immediate beyond the shock front where the magnetic field is more highly ordered, but we did not find an elevated X-ray PD when the X-ray SED peak shifted toward lower energies (from IXPE3-5). %A plausible explanation of the lower PD we see in IXPE4 and 5 might be due to the change of the spectral curvature. The resulting lower synchrotron peak frequency can be attributed to increased electron cooling efficiency, 
The lower \pdx\ we see in IXPE4 and IXPE5 might be attributed to the change of the spectral curvature,
which can be connected to a decrease in the magnetic field strength \cite[e.g.,][]{Abe2024}. Such a decrease would lower the flux and peak frequency of the SED and weaken the radiative energy losses. The former effect increases the spectral curvature, while the latter increases the volume of the emitting region, thereby lowering its degree of polarization.
Alternatively, increased turbulence could also decrease the uniformity of the magnetic field and hence the polarization degree. However, in this case one would not expect the curvature of the SED to increase.
Detailed SED modeling of our observations, planned in a follow-up study, can further elucidate the physical processes in the X-ray emitting region of the jet.

We note that the detected X-ray PD fluctuates between $\sim 10$--$20\%$, far lower than the synchrotron limit \citep[e.g.,][]{Rybicki1979}. This implies that the observed X-ray emission originates from at least a partially turbulent emission region that is close to the site of particle acceleration. This shock-plus-turbulence scenario naturally explains why the elevated PD during IXPE6 was only seen in X-rays. 
We also note the lower ${\rm PD}_{\rm opt}$ values measured in IXPE4 and IXPE5 are still within the $3\sigma$ uncertainty of the average among all six observations, or of the 2022 IXPE observations with higher PD. 
A variety of physical mechanisms could drive the limited fluctuations we see among six IXPE observations. However, based on the overall trend of the elevated PD in the X-rays compared to longer wavelengths, we argue that these six IXPE observations are generally consistent with the energy-stratified shock scenario with a highly turbulent magnetic field structure beyond the shock front.

\section{Conclusion}\label{sec:conclusions}
We have presented three additional IXPE observations of the archetypal high-synchrotron-peaked blazar Mrk~501 obtained in 2023. Combined with the three 2022 observations, our dataset constitutes the first long-term X-ray polarization light curve from a blazar. All of our observations were supplemented with simultaneous multiwavelength campaigns. During the 14-month span of our observations, the source was in an average-to-quiescent flux state across different wavelengths, as well as in a typical radio, infrared, and optical polarization state. 
For Mrk 501, X-ray PD values were found to be generally higher than those in other wavelengths, with occasional drops in X-ray PD to the level consistent with the optical PD. 
The higher X-ray PD measurements compared to those in other wavelengths are common among IXPE-observed HSP blazars 
\citep[e.g.,][]{DiGesu2022-Mrk421,Middei2023,Ehlert2023,Errando2024,kim2024}, as are the fluctuations in the X-ray to optical PD ratio \citep{Errando2024}. These provide further evidence for shock-accelerated electron populations that become energy-stratified as they advect downstream from the shock front inside a turbulent plasma. 
For Mrk 501, the polarization angle at all wavelengths fluctuates only modestly around the jet axis on the plane of the sky, even when the polarization degree changes by a factor of two. This is also common among IXPE observed HSP blazars, implying that the physics of HSP blazars is essentially the same throughout the subclass of AGN, even though Mrk~421 has shown large rotations of the polarization angle \citep{DiGesu2023}. Given such interesting variability patterns,
%although with Mrk~421 being an exception \citep{DiGesu2023}. Given that exception, 
it is important to continue to follow the X-ray and  multi-wavelength polarization, as well as the SED, of Mrk~501 via long-term  monitoring in order to determine whether the consistency of the magnetic field geometry and its relation to particle acceleration remains stable or varies over a longer time scale than sampled thus far. It is also important to continue to develop theories and simulations to explore further the properties of the shock-plus-turbulence model, and perhaps to find other physical scenarios that can reproduce the multiwavelength polarization properties reported here.

\vspace{5mm}
\facilities{Calar Alto, Effesberg-100m, IRAM-30m, IXPE, KANATA, KVN,  LX-200, NOT, NuSTAR, Perkins-1.8m, SMA, SNO, Swift(XRT and UVOT), T60, XMM-Newton}

%% Similar to \facility{}, there is the optional \software command to allow 
%% authors a place to specify which programs were used during the creation of 
%% the manuscript. Authors should list each code and include either a
%% citation or url to the code inside ()s when available.

\software{astropy \citep{2013A&A...558A..33A,2018AJ....156..123A},
ixpeobssim,
xspec
          }

%% Appendix material should be preceded with a single \appendix command.
%% There should be a \section command for each appendix. Mark appendix
%% subsections with the same markup you use in the main body of the paper.

%% Each Appendix (indicated with \section) will be lettered A, B, C, etc.
%% The equation counter will reset when it encounters the \appendix
%% command and will number appendix equations (A1), (A2), etc. The
%% Figure and Table counter will not reset.

% using a non-numbered section to avoid cutoff...
\section*{Acknowledgements}
%\begin{acknowledgements}

The Imaging X-ray Polarimetry Explorer (IXPE) is a joint US and Italian mission.  The US contribution is supported by the National Aeronautics and Space Administration (NASA) and led and managed by its Marshall Space Flight Center (MSFC), with industry partner Ball Aerospace (contract NNM15AA18C)---now, BAE Systems.  The Italian contribution is supported by the Italian Space Agency (Agenzia Spaziale Italiana, ASI) through contract ASI-OHBI-2022-13-I.0, agreements ASI-INAF-2022-19-HH.0 and ASI-INFN-2017.13-H0, and its Space Science Data Center (SSDC) with agreements ASI-INAF-2022-14-HH.0 and ASI-INFN 2021-43-HH.0, and by the Istituto Nazionale di Astrofisica (INAF) and the Istituto Nazionale di Fisica Nucleare (INFN) in Italy.  This research used data products provided by the IXPE Team (MSFC, SSDC, INAF, and INFN) and distributed with additional software tools by the High-Energy Astrophysics Science Archive Research Center (HEASARC), at NASA Goddard Space Flight Center (GSFC). The IAA-CSIC group acknowledges financial support from the grant CEX2021-001131-S funded by MCIN/AEI/10.13039/501100011033 to the Instituto de Astrof\'isica de Andaluc\'ia-CSIC and through grant PID2019-107847RB-C44. The POLAMI observations were carried out at the IRAM 30m Telescope. IRAM is supported by INSU/CNRS (France), MPG (Germany), and IGN (Spain). The Submillimeter Array is a joint project between the Smithsonian Astrophysical Observatory and the Academia Sinica Institute of Astronomy and Astrophysics and is funded by the Smithsonian Institution and the Academia Sinica. Maunakea, the location of the SMA, is a culturally important site for the indigenous Hawaiian people; we are privileged to study the cosmos from its summit. Some of the data reported here are based on observations made with the Nordic Optical Telescope, owned in collaboration with the University of Turku and Aarhus University, and operated jointly by Aarhus University, the University of Turku, and the University of Oslo, representing Denmark, Finland, and Norway, the University of Iceland and Stockholm University at the Observatorio del Roque de los Muchachos, La Palma, Spain, of the Instituto de Astrofisica de Canarias. E.L. was supported by Academy of Finland projects 317636 and 320045. The data presented here were obtained in part with ALFOSC, which is provided by the Instituto de Astrofisica de Andalucia (IAA) under a joint agreement with the University of Copenhagen and NOT. Part of the French contributions is supported by the Scientific Research National Center (CNRS) and the French spatial agency (CNES). 
The USRA coauthors gratefully acknowledge NASA funding through contract 80NSSC24M0035.
The research at Boston University was supported in part by National Science Foundation grant AST-2108622, NASA Fermi Guest Investigator grants 80NSSC23K1507 and 80NSSC22K1571, and NASA Swift Guest Investigator grant 80NSSC22K0537. This study was based in part on observations conducted using the 1.8m Perkins Telescope Observatory (PTO) in Arizona, which is owned and operated by Boston University. This work was supported by JST, the establishment of university fellowships towards the creation of science and technology innovation, Grant Number JPMJFS2129. This work was supported by Japan Society for the Promotion of Science (JSPS) KAKENHI Grant Numbers JP21H01137. This work was also partially supported by the Optical and Near-Infrared Astronomy Inter-University Cooperation Program from the Ministry of Education, Culture, Sports, Science and Technology (MEXT) of Japan. We are grateful to the observation and operating members of the Kanata Telescope. Some of the data are based on observations collected at the Observatorio de Sierra Nevada, owned and operated by the Instituto de Astrof\'{i}sica de Andaluc\'{i}a (IAA-CSIC). Further data are based on observations collected at the Centro Astron\'{o}mico Hispano en Andalucía (CAHA), operated jointly by Junta de Andaluc\'{i}a and Consejo Superior de Investigaciones Cient\'{i}ficas (IAA-CSIC). CC acknowledges support from the European Research Council (ERC) under the HORIZON ERC Grants 2021 program under grant agreement No. 101040021. This work was supported by NSF grant AST-2109127. We acknowledge the use of public data from the Swift data archive. Based on observations obtained with XMM-Newton, an ESA science mission with instruments and contributions directly funded by ESA Member States and NASA. Data from the Steward Observatory spectropolarimetric monitoring project were used. This program was supported by Fermi Guest Investigator grants NNX08AW56G, NNX09AU10G, NNX12AO93G, and NNX15AU81G. We acknowledge funding to support our NOT observations from the Academy of Finland grant  306531. The Very Long Baseline Array is an instrument of the National Radio Astronomy Observatory. The National Radio Astronomy Observatory is a facility of the National Science Foundation operated under a cooperative agreement by Associated Universities, Inc.  S. Kang, S.-S. Lee, W. Y. Cheong, S.-H. Kim, and H.-W. Jeong  were supported by the National Research Foundation of Korea (NRF) grant funded by the Korea government (MIST) (2020R1A2C2009003). The KVN is a facility operated by the Korea Astronomy and Space Science Institute. The KVN operations are supported by KREONET (Korea Research Environment Open NETwork) which is managed and operated by KISTI (Korea Institute of Science and Technology Information). Partly based on observations with the 100-m telescope of the MPIfR (Max-Planck-Institut f\"ur Radioastronomie) at Effelsberg. Observations with the 100-m radio telescope at Effelsberg have received funding from the European Union’s Horizon 2020 research and innovation programme under grant agreement No 101004719 (ORP). The Dipol-2 polarimeter was built in cooperation by the University of Turku, Finland, and the Leibniz Institut f\"{u}r Sonnenphysik, Germany, with support from the Leibniz Association grant SAW-2011-KIS-7. I.L. was supported by the NASA Postdoctoral Program at the Marshall Space Flight Center, administered by Oak Ridge Associated Universities under contract with NASA.
%\end{acknowledgements}

\appendix
\setcounter{figure}{0}
\renewcommand{\thefigure}{A\arabic{figure}}
\section{Supplementary figures for X-ray and multiwavelength data}
\begin{figure*}
     \plottwo{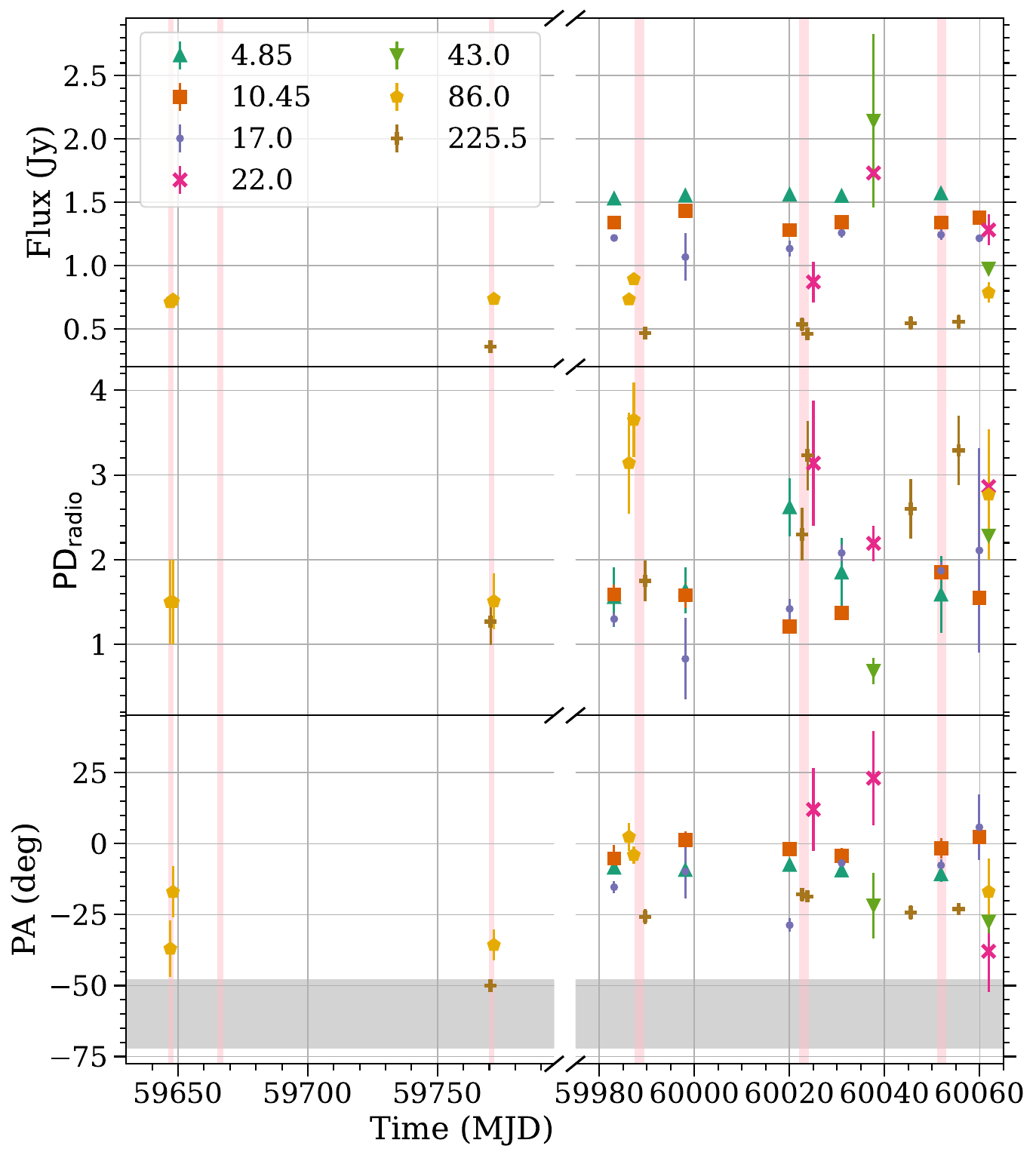}{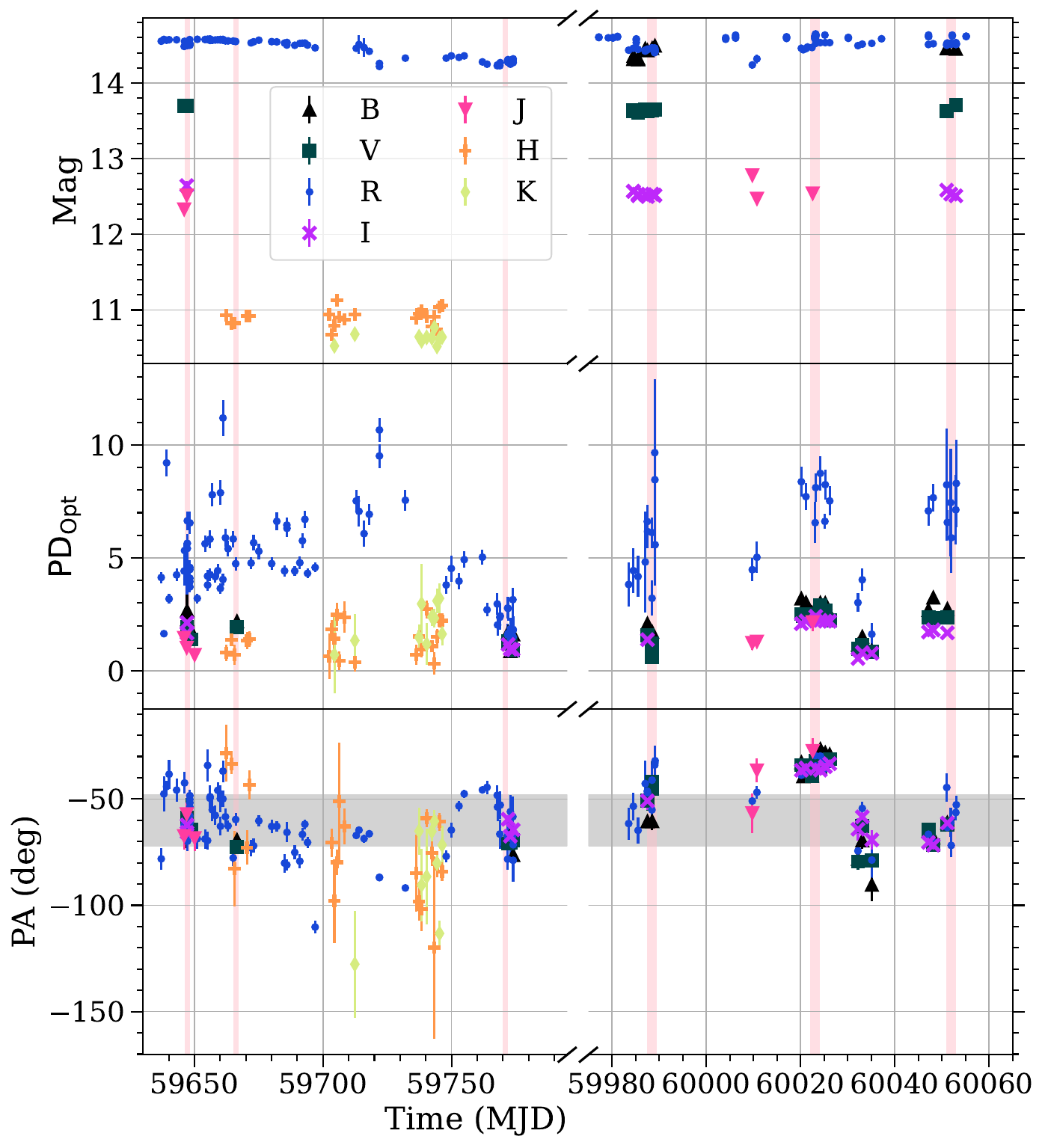}
    \caption{
    Multiwavelength photometric and polarimetric measurements as a function of time. The pink regions mark the duration of the three IXPE observations. The error bars indicate the 68\% confidence (1$\sigma$) uncertainty. The jet axis is marked as the gray region in the bottom panels.
    {\it Left}: Radio observations of Mrk 501 in 2023, observations at different frequencies (in GHz) are marked with different symbols shown in the figure legend. The top panel shows the flux density in Janskys, the middle panel the degree of polarization, and the bottom panel the polarization angle. The error bars indicate the 68\% confidence (1$\sigma$) uncertainty.
    {\it Right}: Optical and infrared observations of Mrk 501 in 2023. The top panel shows the brightness in magnitudes, the middle panel the degree of polarization (in \%), and the bottom panel the polarization angle. 
    }
    \label{fig:radio_obs}
\end{figure*}

\begin{figure*}
%\DeclareGraphicsExtensions{png}
\plottwo{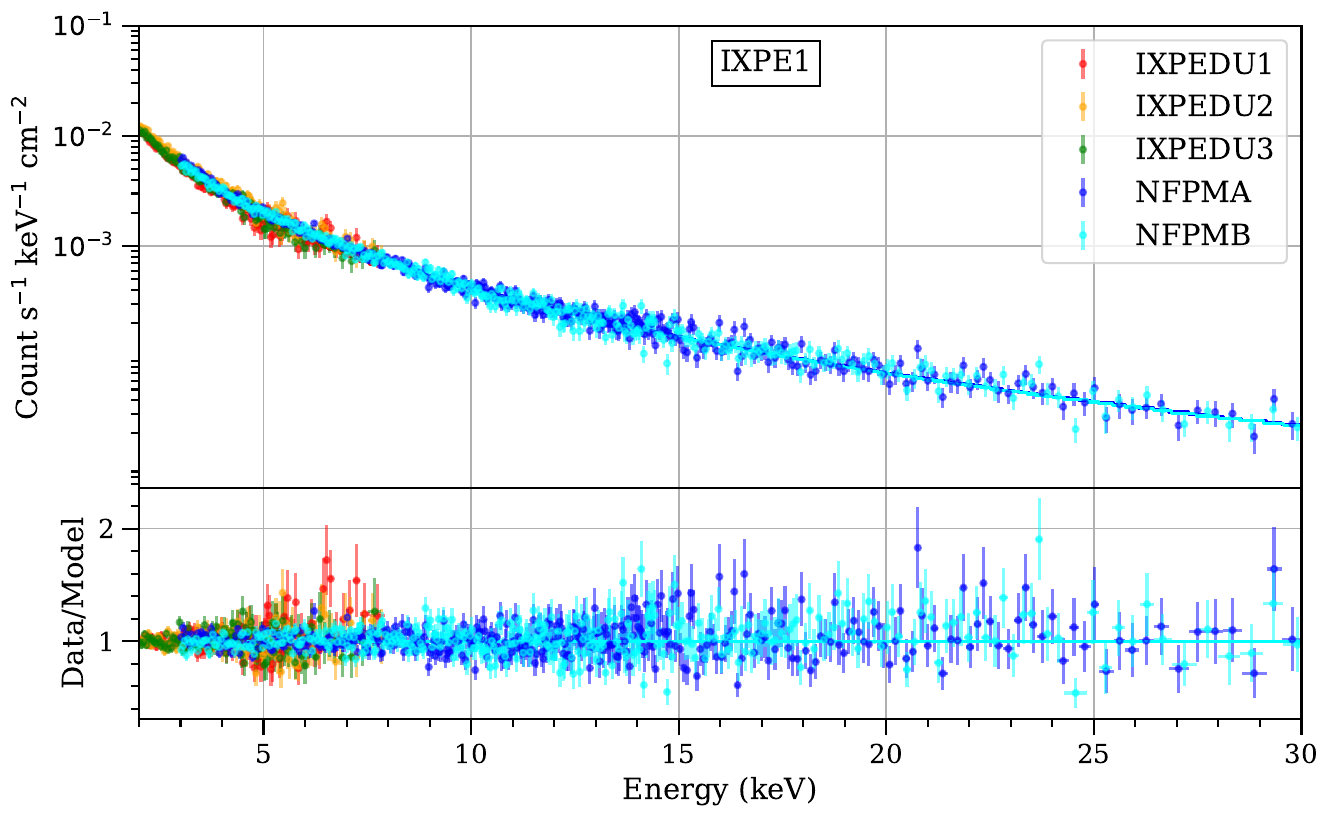}{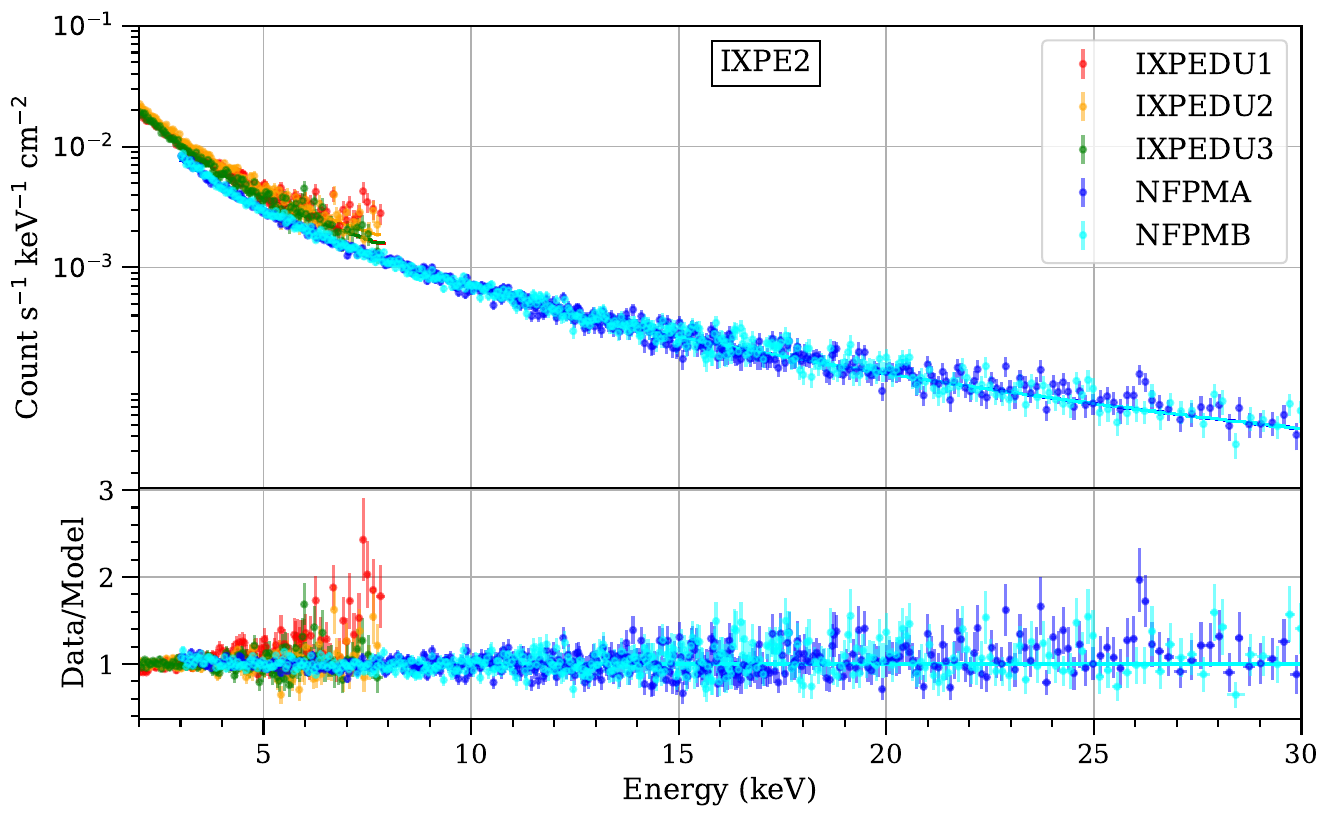}
\plottwo{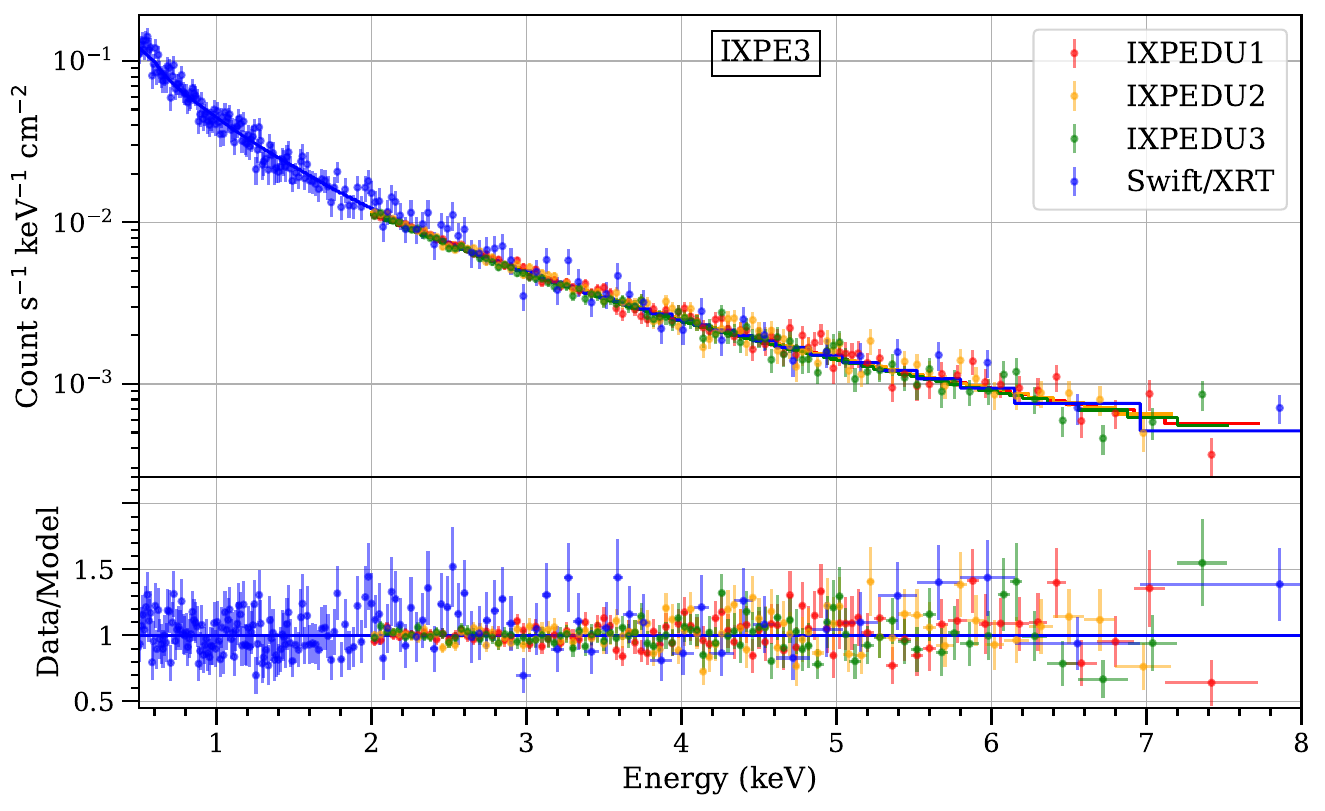}{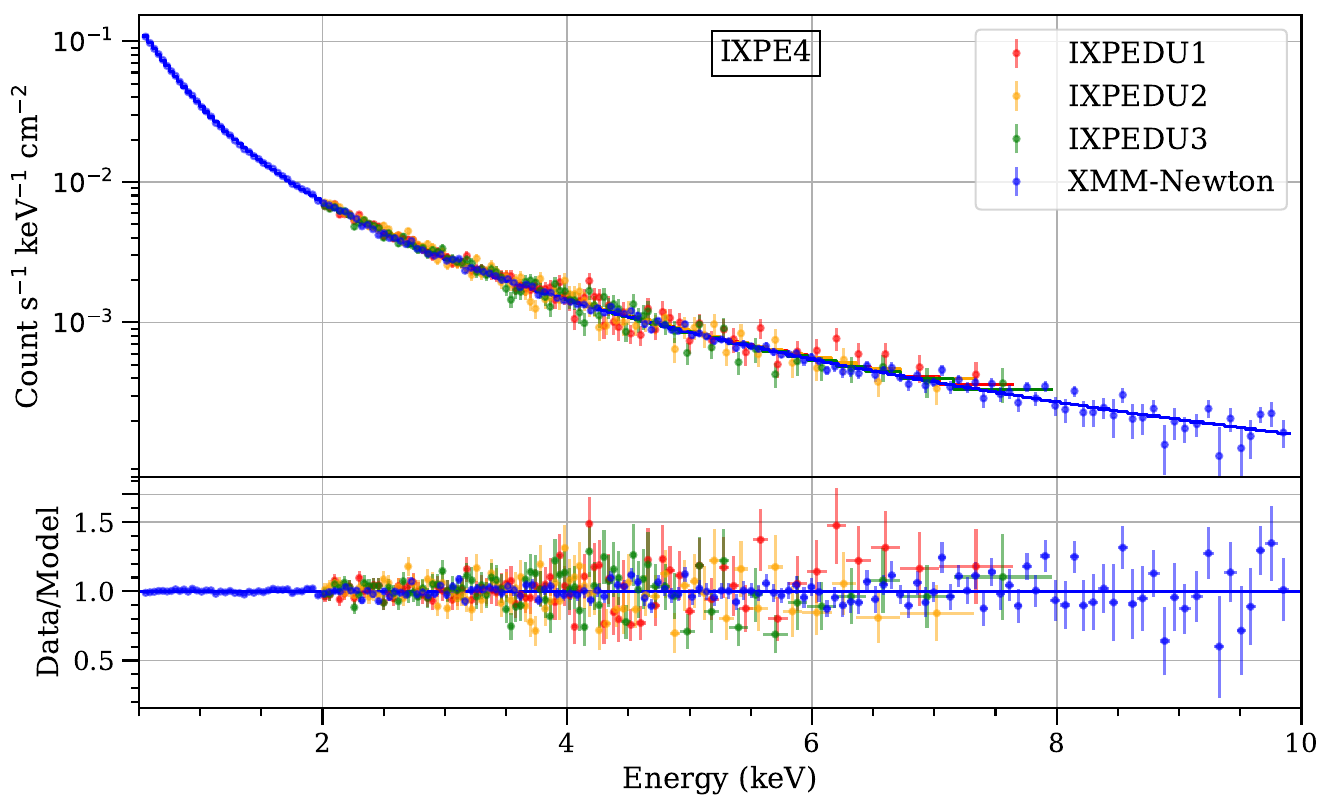}
\plottwo{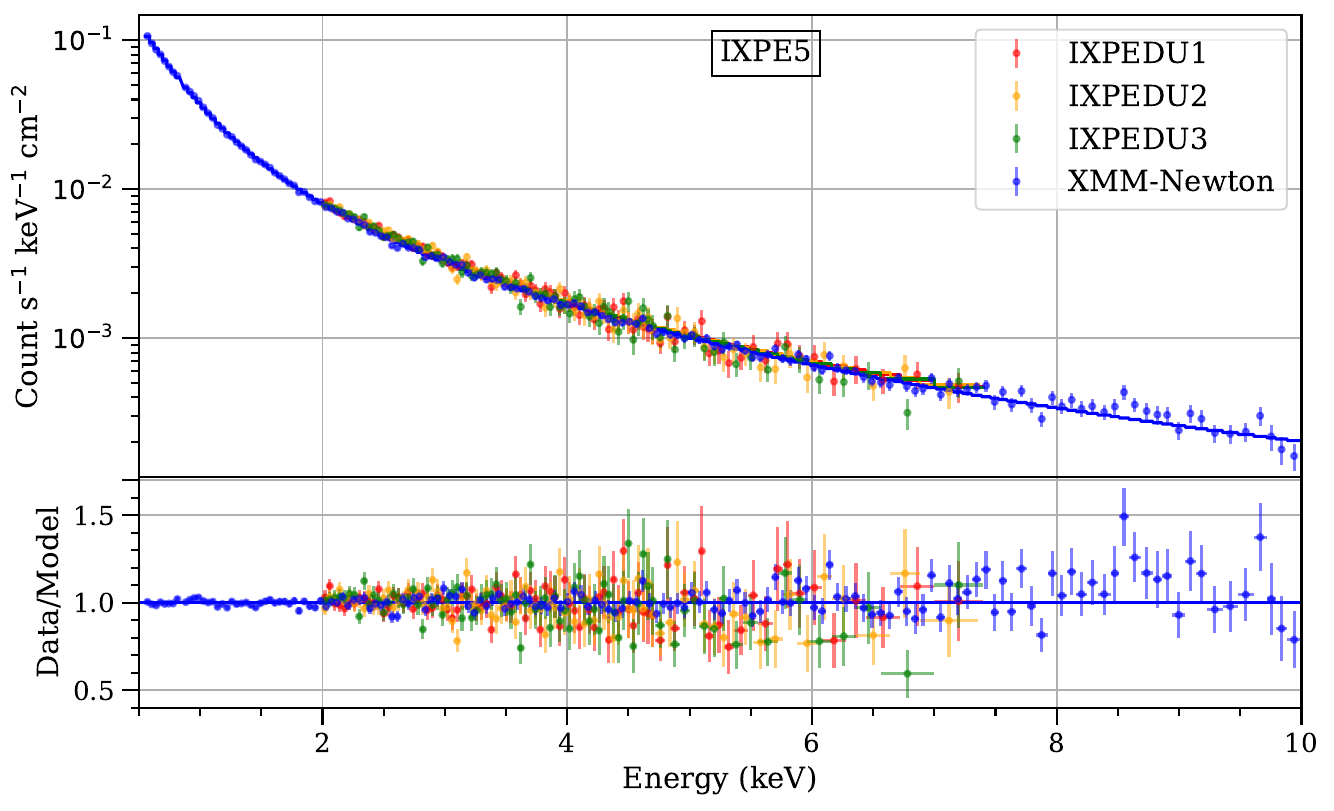}{figs/02004701_xspec.pdf}
\caption{X-ray spectra and best-fit model for each IXPE Mrk~501 observations. Shown here are the IXPE Stokes $I$ spectra and ancillary X-ray data. Different instruments are color coded, with the data and model ratio shown in the bottom panel for each plot. See Section~\ref{sec:specP} for details.}
\label{fig:appspecplots}
\end{figure*}

\begin{figure*}
\plottwo{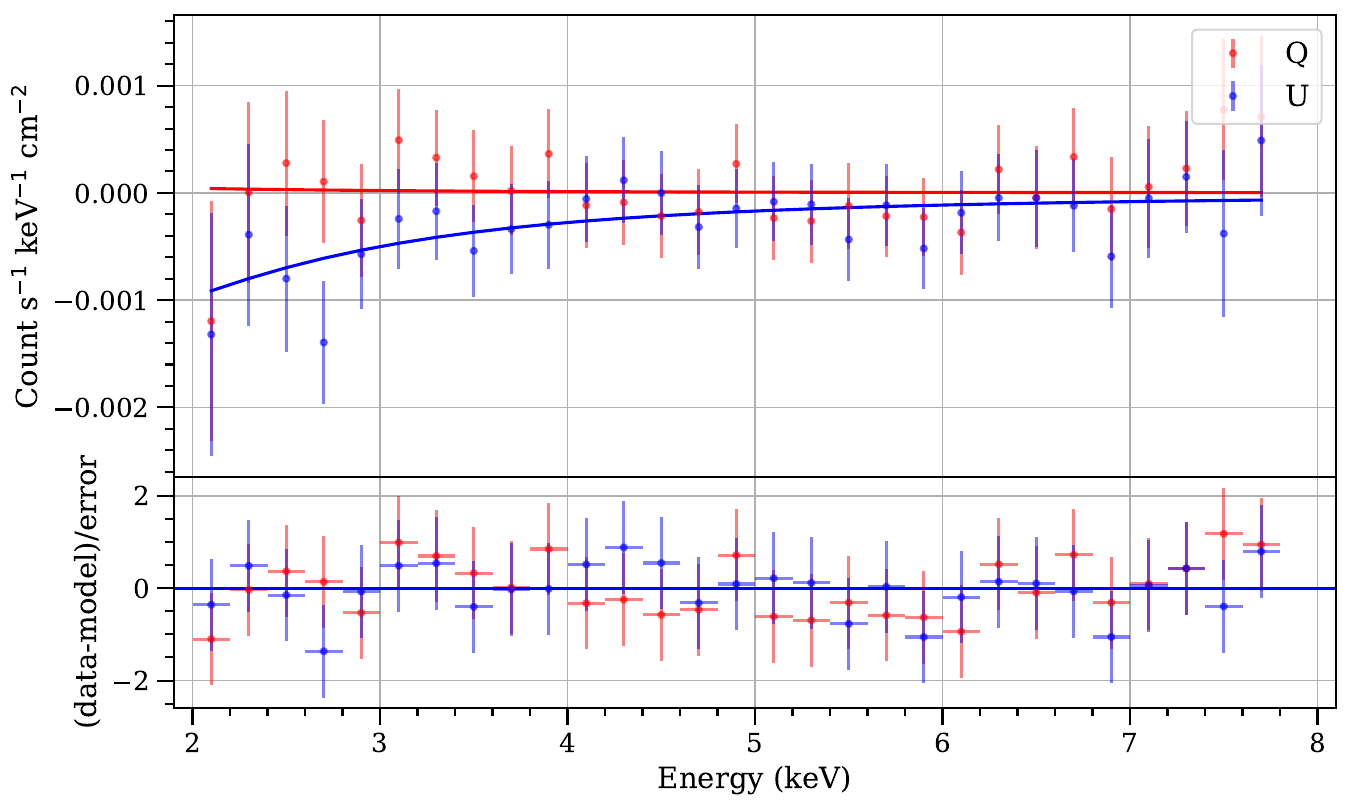}{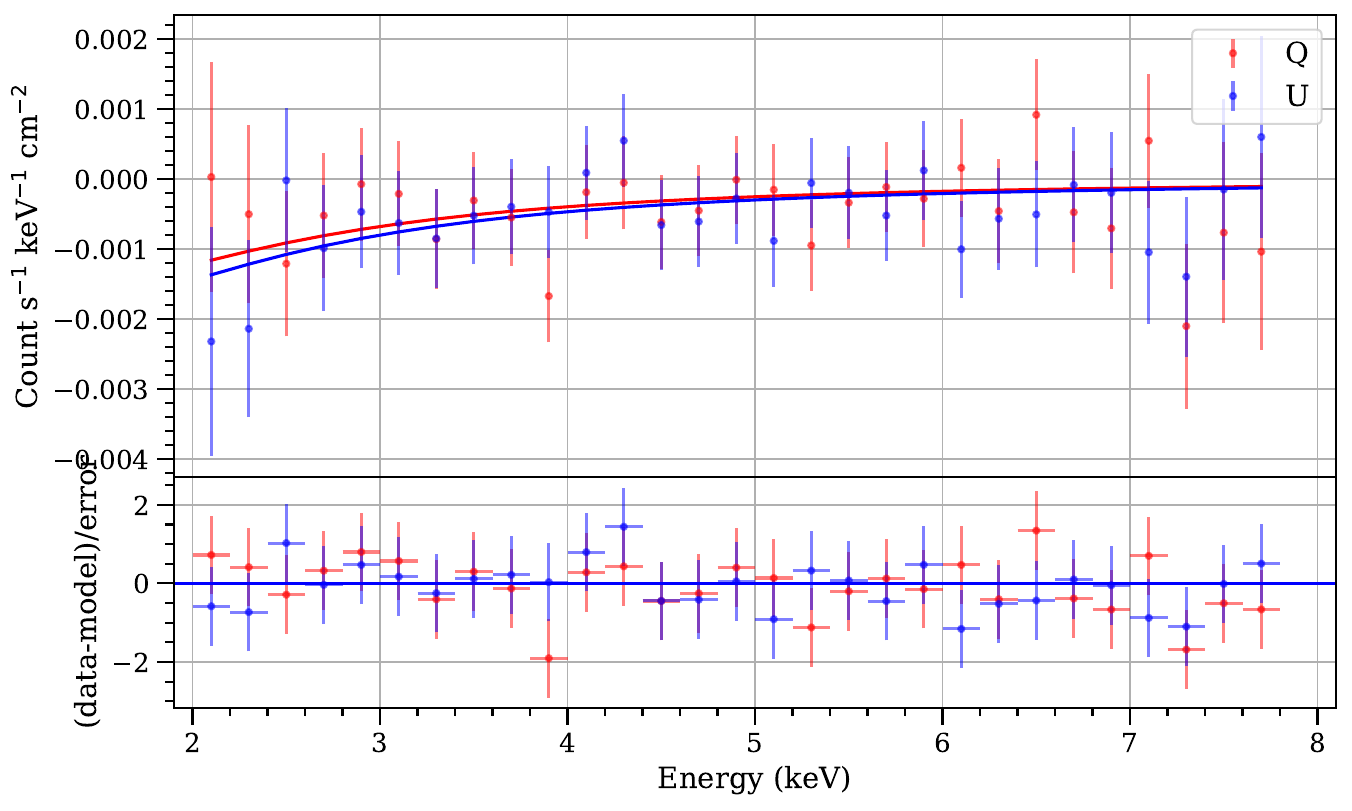}
\plottwo{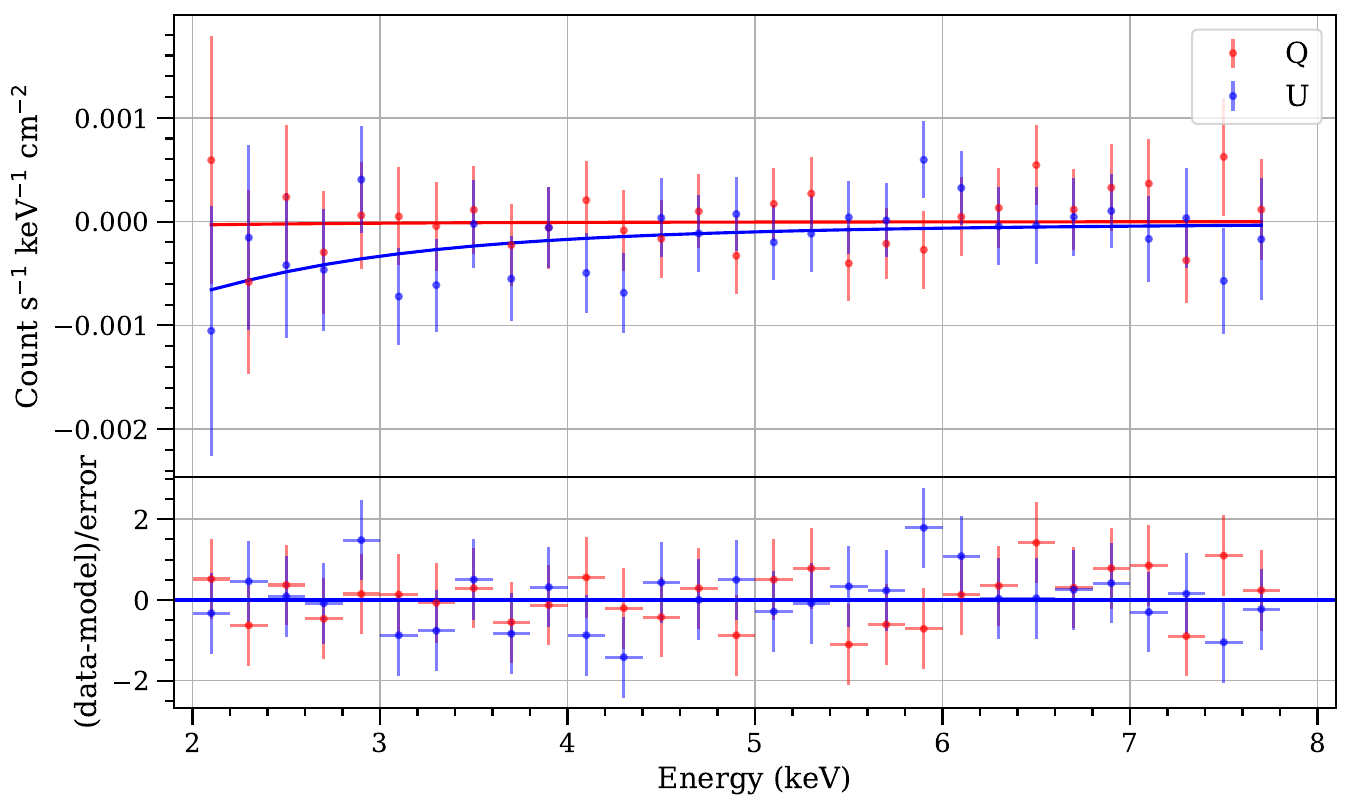}{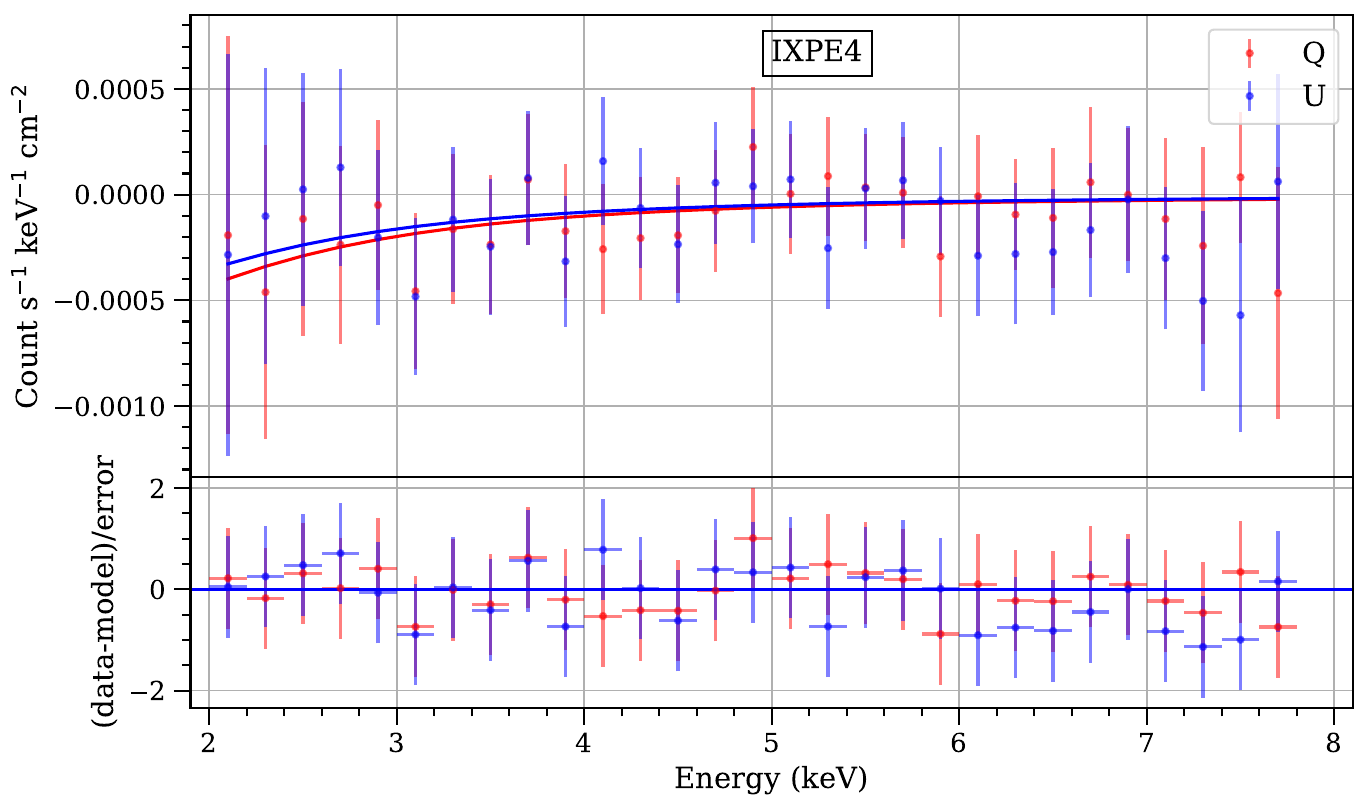}
\plottwo{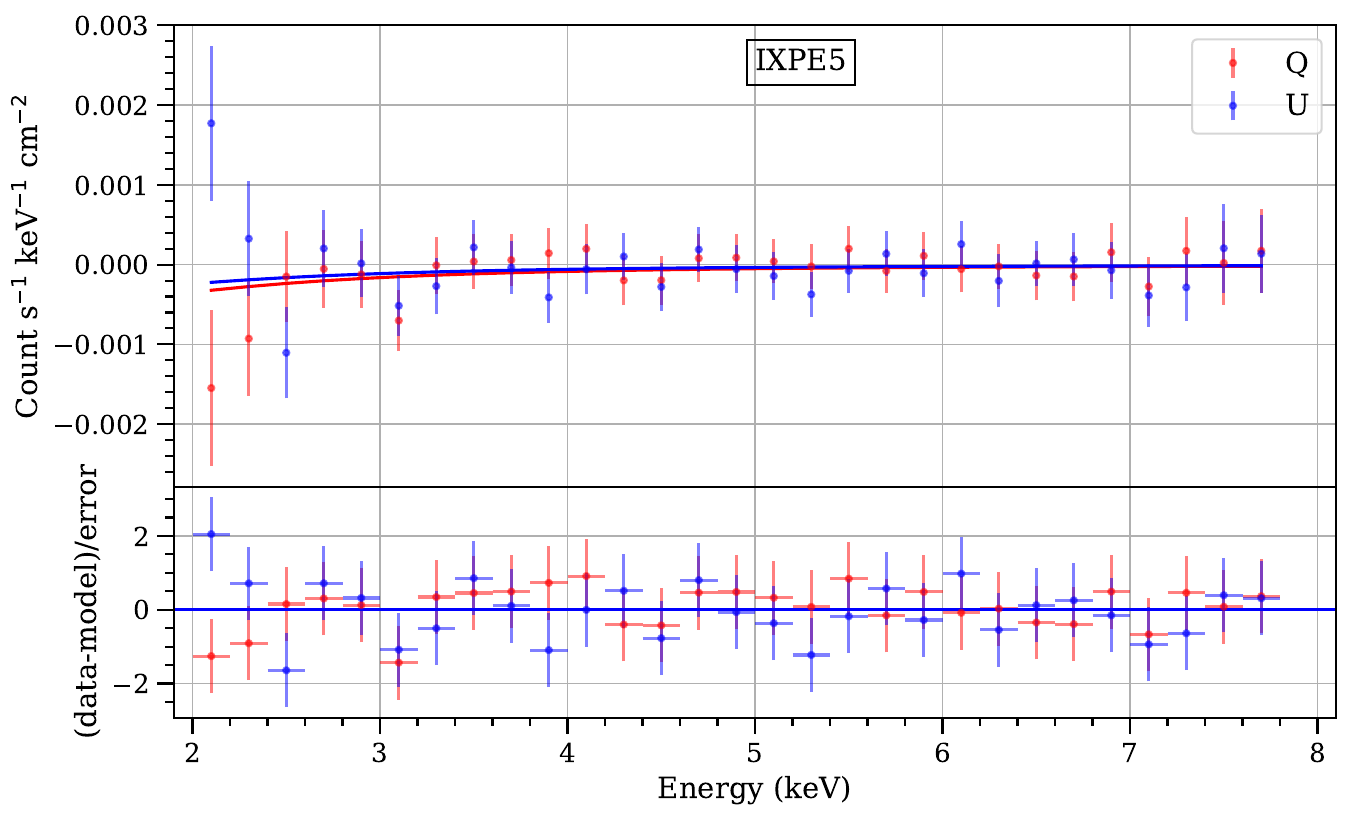}{figs/02004701_xspec_QU.pdf}
\caption{Similar to Figure~\ref{fig:appspecplots}, but only showing the IXPE Stokes $Q$ and $U$ spectra.}
\label{fig:appquplots}
\end{figure*}

We show the complete multiwavelength data in the radio and optical-IR bands in Figure~\ref{fig:radio_obs}. 
Spectro-polarimetric data and best-fit models for individual IXPE observations and their ancillary data are also shown in Figure~\ref{fig:appspecplots} and Figure~\ref{fig:appquplots}. 

As shown in Fig.~\ref{fig:radio_obs}, some IXPE observations were covered with multiple optical measurements. Specifically, there are 14, 1, 4, 7, 2, and 5 R-band observations from IXPE1 to IXPE6, respectively. We did not see substantial variations in the optical PD and PA. During IXPE4, PD$_{\rm opt}$ appears to have increased. However, within the IXPE4 measurement, the observation with the most significant deviation from the average value also exhibited large uncertainties, placing it only $1.6\sigma$ away from the average. 
For the radio data, no single IXPE observation included more than one radio measurement at the same frequency. IXPE4 and IXPE5 do have additional radio measurements immediately before or after the IXPE observations. However, they are within 1$\sigma$ uncertainty of the radio measurements contemporaneous to the IXPE observations. The lack of apparent correlations between different wavelengths is still consistent with the energy-stratified shock scenario in which emission at different wavelengths is thought to originate from different regions.

\bibliographystyle{aasjournal}

\end{document}

%% file: authorlist.tex
%the author list is under construction, please, do not modify it 
% Tier 1a - contributed in analysis 
% 

\author[0000-0002-4945-5079]{Chien-Ting J. Chen}
\affiliation{Science and Technology Institute, Universities Space Research Association, Huntsville, AL 35805, USA}
\affiliation{Astrophysics Office, NASA Marshall Space Flight Center, ST12, Huntsville, AL 35812, USA}
%email: ct.chen@nasa.gov

\author[0000-0001-9200-4006]{Ioannis Liodakis}
\affiliation{NASA Marshall Space Flight Center, Huntsville, AL 35812, USA}
\affiliation{Institute of Astrophysics, Foundation for Research and Technology-Hellas, GR-70013 Heraklion, Greece}
%email:yannis.liodakis@gmail.com

\author[0000-0001-9815-9092]{Riccardo Middei}
\affiliation{Space Science Data Center, Agenzia Spaziale Italiana, Via del Politecnico snc, 00133 Roma, Italy}
\affiliation{INAF Osservatorio Astronomico di Roma, Via Frascati 33, 00040 Monte Porzio Catone (RM), Italy}
%email:riccardo.middei@ssdc.asi.it

\author[0000-0001-5717-3736]{Dawoon E. Kim}
\affiliation{INAF Istituto di Astrofisica e Planetologia Spaziali, Via del Fosso del Cavaliere 100, 00133 Roma, Italy}
\affiliation{Dipartimento di Fisica, Universit\`a degli Studi di Roma ``La Sapienza'', Piazzale Aldo Moro 5, 00185 Roma, Italy}
\affiliation{Dipartimento di Fisica, Universit\`a degli Studi di Roma ``Tor Vergata'', Via della Ricerca Scientifica 1, 00133 Roma, Italy}
%email:dawoon.kim@inaf.it

% Tier 1b - participated in discussion/editing/analysis/interpretation

\author[0000-0002-5614-5028]{Laura Di Gesu}
\affiliation{ASI - Agenzia Spaziale Italiana, Via del Politecnico snc, 00133 Roma, Italy}
%email: laura.digesu@est.asi.it

\author[0000-0003-0331-3259]{Alessandro Di Marco}
\affiliation{INAF Istituto di Astrofisica e Planetologia Spaziali, Via del Fosso del Cavaliere 100, 00133 Roma, Italy}
%email:alessandro.dimarco@inaf.it

\author[0000-0003-4420-2838]{Steven R. Ehlert}
\affiliation{NASA Marshall Space Flight Center, Huntsville, AL 35812, USA}
%email:steven.r.ehlert@nasa.gov

\author[0000-0002-1853-863X]{Manel Errando}
\affiliation{Physics Department and McDonnell Center for the Space Sciences, Washington University in St. Louis, St. Louis, MO 63130, USA}
%email:errando@wustl.edu
\author[0000-0002-6548-5622]{Michela Negro} 
\affiliation{Department of Physics and Astronomy, Louisiana State University, Baton Rouge, LA 70803, USA}
%michelanegro@lsu.edu

\author[0000-0001-6158-1708]{Svetlana G. Jorstad}
\affiliation{Institute for Astrophysical Research, Boston University, 725 Commonwealth Avenue, Boston, MA 02215, USA}
\affiliation{Saint Petersburg State University, 7/9 Universitetskaya nab., 199034 St. Petersburg, Russia}
%email: jorstad@bu.edu

% ...
%email: s.zane@ucl.ac.uk